\documentclass[journal]{IEEEtran}
\usepackage{amsmath,amsfonts}
\usepackage{algorithmic}
\usepackage{algorithm}
\usepackage{array}
\usepackage{multirow}
\usepackage[caption=false,font=normalsize,labelfont=sf,textfont=sf]{subfig}
\usepackage{textcomp}
\usepackage{stfloats}
\usepackage{url}
\usepackage{verbatim}
\usepackage{graphicx}
\usepackage{cite}
\usepackage{epstopdf}
\usepackage{color}
\usepackage{threeparttable}
\usepackage{diagbox}
\usepackage{amssymb}
\usepackage{makecell}
\usepackage{hyperref}

\begin{document}
\title{Index Modulation for Fluid Antenna-Assisted MIMO Communications: System Design and Performance Analysis}
\author{Jing Zhu, \textit{Student Member, IEEE}, Gaojie Chen, \textit{Senior Member, IEEE}, Pengyu Gao, Pei Xiao, \textit{Senior Member, IEEE}, Zihuai Lin, \textit{Senior Member, IEEE}, and Atta Quddus
\thanks{
J. Zhu, G. Chen, P. Gao, P. Xiao and A. Quddus are with 5G and 6G Innovation centre, Institute for Communication Systems (ICS) of University of Surrey, Guildford, GU2 7XH, UK (e-mail: \{j.zhu, gaojie.chen, p.gao, p.xiao, a.quddus\}@surrey.ac.uk).

Z. Lin is with the School of Electrical and Information Engineering, University of Sydney, Sydney, NSW 2006, Australia (e-mail: zihuai.lin@sydney.edu.au).
}

}

\maketitle

\begin{abstract}
In this paper, we propose a transmission mechanism for fluid antennas (FAs) enabled multiple-input multiple-output (MIMO) communication systems based on index modulation (IM), named FA-IM, which incorporates the principle of IM into FAs-assisted MIMO system to improve the spectral efficiency (SE) without increasing the hardware complexity. In FA-IM, the information bits are mapped not only to the modulation symbols, but also the index of FA position patterns. Additionally, the FA position pattern codebook is carefully designed to further enhance the system performance by maximizing the effective channel gains. Then, a low-complexity detector, referred to efficient sparse Bayesian detector, is proposed by exploiting the inherent sparsity of the transmitted FA-IM signal vectors. Finally, a closed-form expression for the upper bound on the average bit error probability (ABEP) is derived under the finite-path and infinite-path channel condition. Simulation results show that the proposed scheme is capable of improving the SE performance compared to the existing FAs-assisted MIMO and the fixed position antennas (FPAs)-assisted MIMO systems while obviating any additional hardware costs. It has also been shown that the proposed scheme outperforms the conventional FA-assisted MIMO scheme in terms of error performance under the same transmission rate.
\end{abstract}


\IEEEpeerreviewmaketitle

\section{Introduction}
\IEEEPARstart{T}{he} rapid progress in multiple-input multiple-output (MIMO) technologies has ushered in a tranformative era for wireless communication systems, significantly amplifying their capacity by harnessing the untapped degrees of freedom (DoFs) in the spatial domain \cite{bjornson2016massive,albreem2019massive,wang2019overview}. By leveraging the principles of MIMO, communication systems can now deploy multiple antennas both at the transmitter and receiver, thereby creating multiple spatial streams for data transmissions. This spatial diversity, facilitated by the varying propagation paths between antennas, offers a wealth of independent DoFs, fundamentally altering the traditional constraints on communication capacity. However, in conventional MIMO systems, it is customary to employ fixed position antennas (FPAs) with inter-antenna spacing typically no smaller than half a wavelength at the transceivers. As a result, the FPAs-assisted MIMO imposes limitation on the exploitation of diversity and spatial multiplexing since it cannot explore the full spatial variation of wireless channels within a given spatial filed. 

To solve this problem, fluid antenna system (FAS) was proposed in \cite{wong2020fluid}, where the fluid antenna has the capability to move to any of the potential positions within a set linear range in order to capture the most robust signal. Subsequently, the authors of \cite{wong2021fluid} extended FAS to multiuser scenarios, namely fluid antenna multiple access (FAMA) system, which can concurrently support transceivers operating under distinct channel conditions. In FAMA systems, interference in deep fading regions can be effectively mitigated by optimizing the positions of users' fluid antennas, thereby providing favorable channel conditions for desired signals. The two-dimensional (2D) or three-dimensional (3D) design of FAS was introduced in \cite{wong2022bruce,shen2022radiation}. As another 2D/3D implementation of FAS, movable antennas (MAs) system was proposed in \cite{zhu2022modeling}, where the conventional solid antenna can be freely moved in the 2D/3D space by employing mechanical drives. %However, due to the constraints imposed by physical characteristics, liquid antennas are only capable of moving within a one-dimensional (1D) space and it is hard to form antenna arrays. In contrast, movable antennas (MAs) system was proposed in \cite{zhu2022modeling}, where the conventional solid antenna can be freely moved in the two-dimensional (2D) or three-dimensional (3D) space by employing mechanical drives.

MAs can be equipped in MIMO systems to exploit the full spatial diversity by varying the antenna position in a given finite region at the transceiver, which can be regarded as a promising technology in future wireless communications, especially in machine-type communications (MTC) \cite{beyene2017performance,dowhuszko2016performance}. It is worth noting that the MAs based on mechanical movement is impractical in high-mobility scenarios, due to the significant time required for antenna shifting. The delayed response hinders its  adaptation to the dynamic channel conditions, posing a primary challenge for MAs in high-mobility communications. In MTC, the Internet of Things (IoT) devices are deployed in limited areas at fixed positions, where the surrounding environment may change over time, leading to slowly-varying wireless channels. In such low-mobility scenarios, narrow-band MTC possess limited time and frequency diversity to enhance transmission reliability. Therefore, compared to conventional FPAs, MA becomes a viable technique for achieving higher spatial diversity gains for slowly varying channels.  In \cite{zhu2022modeling}, a channel model based on field response was introduced for the single-MA system. This model analyzed the signal-to-noise ratio (SNR) improvement over its FPA counterpart, considering both deterministic and stochastic channel scenarios. The authors of \cite{ma2022mimo} applied MA into MIMO systems, where the maximum channel capacity is achieved by jointly optimizing the positions of transmitter and receiver MAs. It was also verified that the MA-assisted MIMO system outperforms the conventional FPA-assisted MIMO in terms of channel capacity. In \cite{zhu2023movable}, the architecture, channel characterization of MA were introduced, followed by the major performance gain analysis and typical applications and technique challenges discussions. Channel estimation of MA-assisted MIMO was considered in \cite{ma2023compressed}, where compressed sensing (CS) based method is employed to efficiently solve the channel estimation problem by exploiting the sparse representation of the filed-response based channel model. The above mentioned MA systems \cite{zhu2022modeling, ma2022mimo, zhu2023movable, ma2023compressed} assume that the positions of MA elements can be freely adjusted  within a given continuous region to attain full spatial diversity, requiring the use of electromechanical devices with infinite resolution. Another approach to MA implementation has been presented in \cite{zhuravlev2015experimental,basbug2017design,wu2023movable}, where the motion of MAs is discrete. In this case, the transmitter area is quantified in a pace \cite{basbug2017design}, resulting in finite spatial resolution, albeit at the cost of spatial diversity loss. Furthermore, the MA-assisted MIMO system offers low spectral efficiency (SE) performance when equipped with small-scale antennas, otherwise they still face high hardware cost problem. 

Index modulation (IM) represents a promising solution to reduce the hardware cost by reducing the number of RF chains and at the same time, enhance SE by exploiting the indices of antennas, subcarriers, time slots, channel states to convey additional information \cite{dang2017adaptive,zhu2022index,dang2018lexicographic,wen2019index,wen2021joint}.
IM-aided systems offer distinctive benefits, such as high energy efficiency (EE), low hardware complexity and flexible system structures, by introducing extra dimensions in contrast to conventional modulation systems. The concept of IM has been integrated with conventional FPA-assited MIMO systems. Specifically, spatial modulation (SM) aided MIMOs was proposed in \cite{mesleh2008spatial}, where the active antenna index is exploited as an additional dimension to transmit information. To further enhance the SE and error performance, a set of SM-variants were investigated, such as  generalized SM (GSM) \cite{xiao2014low}, quadrature SM (QSM) \cite{li2016generalized}, enhanced SM (ESM) \cite{yang2019enhanced}, precoding SM (PSM) \cite{zhu2018low} and so on.  More recently, IM has also been widely applied in high-frequency scenarios. In \cite{gao2019spatial}, generalized beamspace modulation (GBM) was proposed in millimeter wave (mmWave) massive MIMO systems, which utilizes the distinctive characteristics of mmWave massive MIMO. Spatial scattering modulation (SSM) was proposed in \cite{ding2017spatial} by indexing a set of orthogonal paths in the mmWave environment. 

%The above IM schemes were employed in FPAs-assisted MIMO systems, which require large-scale antennas to exploit full spatial diversity, thus incurring high hardware cost. In light of the challenges faced by the conventional IM based FPAs-assisted MIMO and MAs-assisted MIMO systems, the incorporation of IM into MAs-assisted MIMO systems emerges as a compelling and intriguing approach due to its potential to enhance the SE performance while reducing the hardware cost. 
%\textcolor{red}{Compared to the MA-assisted MIMO system, the proposed scheme is able to achieve SE enhancement with the aid of IM techniques. On the other hand, it can achieve the same SE as the FPA-IM system, but with much fewer antenna elements.} 

The above IM schemes were employed in FPAs-assisted MIMO systems, termed as FPA-IM, which require large-scale antennas to exploit full spatial diversity, thus incurring high hardware cost. To the best of our knowledge, the integration of IM and FA-assisted MIMO has not been addressed in the literature. In this paper, we propose a transmission mechanism for FA-assisted MIMO system based on IM, named FA-IM, with the primary objective of enhancing SE while  concurrently reducing hardware costs. In comparison to the classic FA-assisted MIMO system \cite{zhu2022modeling}, the proposed scheme achieves SE enhancement through the integration of  IM techniques. Additionally, it matches the SE performance of the FPA-IM system, but requires substantially fewer antenna elements. It is worth noting that although the proposed scheme is mathematically similar to the antenna selection assisted FPA (AS-FPA) scheme \cite{sanayei2004antenna}, there are two key distinctions: 1) the AS-FPA requires a large number of antennas for high diversity, while the proposed FA-IM can acquire the same spatial diversity with much fewer antennas moving in the same region. 2) the AS-FPA aims to reduce the cost and complexity brought by numerous RF chains, without conveying extra information bits. By contrast, the FA-IM enhances SE by introducing index bits without any extra hardware costs.

The main contributions of this paper are summarized as follows.
\begin{itemize}
    \item 
     We design a novel transmission scheme, called FA-IM, by creating a synergy between the IM technique and the FAs-assisted MIMO system, providing significant SE performance enhancement. The information bits of the proposed scheme are mapped not only to the $M$-ary quadrature amplitude modulation (QAM) constellations, but also the index of FA position patterns. Moreover, the FA position pattern codebook is carefully designed by maximizing the effective channel gains. 
     
    \item 
     We propose a low-complexity detector, namely efficient sparse Bayesian detector, which inherently advances the original sparse Bayesian algorithm by restructuring the iteration process. To be specific, we simplify the iteration process by pruning the initial FA position candidate, thus achieving significant computational complexity reduction. 

    \item 
    We analyze the SE performance of the proposed FA-IM, which demonstrates its superiority over its counterparts in FPAs assisted MIMO system and FAs-assisted MIMO without IM. Meanwhile, the closed-form expression for the upper bound on the average bit error probability (ABEP) is derived to validate the benefits of the proposed scheme.

    \item 
    Simulation results show that 1) the theoretical analysis of ABEP is accurate; 2) the bit error rate (BER) performance of the proposed FA-IM with FA position pattern codebook design outperforms the scheme with random FA position pattern codebook; 3) the proposed scheme offers better BER performance and robustness compared to the benchmark systems under the same SE conditions, and 4) the proposed low-complexity efficient sparse Bayesian detector saves 99.9$\%$ and $25\%$ computational cost compared to the optimal maximum likelihood (ML) and the original sparse Bayesian detectors, respectively. 
    
\end{itemize}

The rest of this paper is organized as follows. In Section II, the channel and signal models of the proposed FA-IM system are introduced. Section III presents an efficient sparse Bayesian detector. Then, the ABEP performance analysis is presented in Section IV. This is followed by simulation results in Section V. Finally, Section VI concludes this paper.

\textit{Notations:} Scalar variables are denoted by normal-face letters, while boldface capital and lowercase symbols represent matrices and column vectors, respectively. $\left|  \cdot  \right|$, ${\tbinom{n}{k}} $ and $\left\lfloor  \cdot  \right\rfloor $ refer to the absolute value, the binomial coefficient and the floor operation, respectively. ${\left(  \cdot  \right)^H}$, ${\left(  \cdot  \right)^{ - 1}}$ and ${\left(  \cdot  \right)^{ \dag}}$ denote Hermitian transpose, inverse and pseudo inverse, respectively. $\rm{diag}\left(  \cdot  \right)$ and ${\mathbb{E}}\left(  \cdot  \right)$ stand for the diagonal and expectation operation, respectively. ${\Re} \left\{  \cdot  \right\}$ represents the real part of a complex variable. ${\left\| {\bf{A}} \right\|}$ and ${\rm det}(\bf A)$ denote the Frobenius norm and determinant of ${\bf{A}}$, respectively. ${{\mathbb{C}}^{{M} \times {N}}}$ represents the space of $M \times N$ complex matrices. ${{\bf{I}}_N}$ stands for an $N$-dimensional identity matrix. $\mathbf{1}_L$ refers to an $L$-dimensional vector with all elements equal to 1.

\section{Proposed Fluid Antenna-Assisted Index Modulation}
\begin{figure*}
\centering
\includegraphics[width=0.95\textwidth]{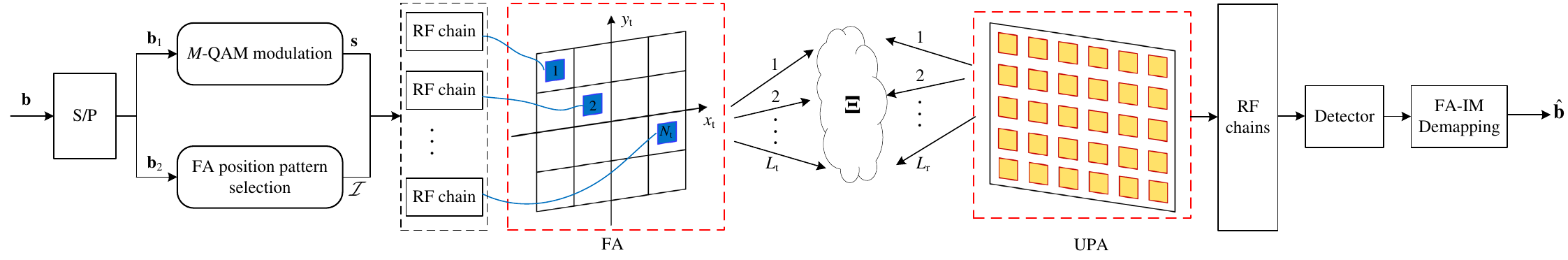}
\caption{Block diagram of the proposed FA-IM system}\label{System_Model}
\end{figure*}

The proposed FA-IM system model is depicted in Fig. \ref{System_Model}. In this 2D configuration, $N_t$ transmit FAs are linked to the RF chains using flexible connectors like coaxial cables\footnote{3D FAs provide additional degrees of freedom for IM, which can further enhance the SE. The system design and performance analysis of 3D FA-IM will be left for our future work.}. Specifically, the whole transmit FA region is divided into $P=P_1 \times P_2$ grids, where $P_1$ and $P_2$ denote the number of grids along horizontal and vertical directions, respectively. According to the property of FA, the positions of the FAs can be altered mechanically using drive components, such as stepper motors. In addition to motor-based FAs, an alternative and efficient approach to implementing FAs is through the microelectromechanical systems (MEMS)-integrated antenna \cite{balanis2008mems}. It has compact components and thus can be easily integrated in small devices.
In this work, we assume that the $N_t$ FAs  can move freely in these $P$ parts.
The receiver (Rx) is equipped with a fixed uniform planar array (UPA) of size $N_r=N_1 \times N_2$, where $N_1$ and $N_2$ represent the number of antennas along horizontal and vertical directions, respectively.

\subsection{Channel model}
Firstly, we establish the Cartesian coordinate systems, ${x_{\rm{t}}} - {O_{\rm{t}}} - {y_{\rm{t}}}$ and ${x_{\rm{r}}} - {O_{\rm{r}}} - {y_{\rm{r}}}$, to describe the positions of the transmit FAs and receive UPA, respectively. The coordinate set of the all the possible transmit FA positions are denoted as ${\cal{T}}=[{\bf{t}}_{1,1},{\bf{t}}_{1,2},...,{\bf{t}}_{m,n},...,{\bf{t}}_{P_1,P_2}]$, where ${{\bf{t}}_{m,n}} = [{x_{{\rm{t,}}m}},{y_{{\rm{t,}}n}}]$, $m = 1,...,{P_1}$, $n = 1,...,{P_2}$. The  coordinate set of the receive UPA elements are denoted as ${\cal{R}}=[{\bf{r}}_{1,1},{\bf{r}}_{1,2},...,{\bf{r}}_{p,q},...,{\bf{r}}_{N_1,N_2}]$ with
${{\bf{r}}_{p,q}} = [{x_{{\rm{r,}}p}},{y_{{\rm{r,}}q}}]$, $p = 1,...,{N_1}$, $q = 1,...,{N_2}$.

For the considered FA-IM  system, the channel response is contingent upon the positions of the antennas. Consequently, the channel coefficient can be expressed as a function of the positions of both the transmitting and receiving antennas, i.e., $h({\bf{t}}_{m,n},{\bf{r}}_{p,q})$.
In this paper, we make the assumption that the area available for antenna movement is significantly smaller than the propagation distance between the Tx and Rx, ensuring that the far-field condition is met at both the Tx and Rx sides \cite{zhu2022modeling}.
Hence, the plane-wave model can be employed to characterize the field response between the transmitting and receiving areas. In simpler terms, the angles of departure (AoDs), angles of arrival (AoAs), and amplitudes of the complex coefficients for the multiple channel paths remain constant regardless of the positions of the FAs. The only variations occur in the phases of the multipath channels within the transmit/receive region.

\begin{figure}
\centering
\includegraphics[width=0.5\textwidth]{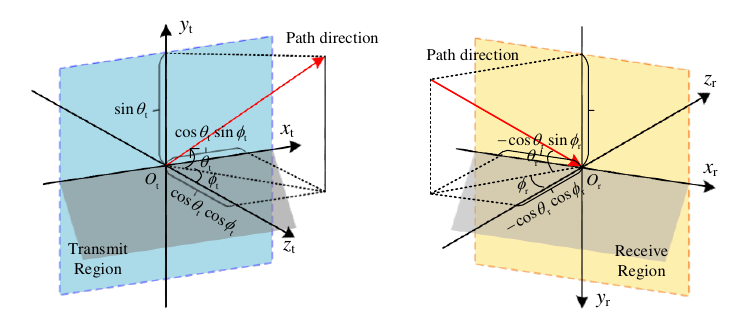}
\caption{Illustration of the coordinates and spatial angles for the transmit and receive regions}\label{Phase_system}
\end{figure}

At the Tx side, we use the notation $L_t$ to represent the number of transmit paths. As depicted in Fig. \ref{Phase_system}, the elevation and azimuth AoDs of the $j$th transmit path are denoted as ${\theta _{{\rm{t}},j}} \in [ - \frac{\pi }{2},\frac{\pi }{2}]$ and ${\phi _{{\rm{t}},j}} \in [ - \frac{\pi }{2},\frac{\pi }{2}]$, $1 \le j \le L_{\rm{t}}$, respectively. At the Rx side, we use the notation $L_r$ to represent the number of receive paths. The elevation and azimuth AoAs of the $i$th receive path are respectively denoted as ${\theta _{{\rm{r}},i}} \in [ - \frac{\pi }{2},\frac{\pi }{2}]$ and ${\phi _{{\rm{r}},i}} \in [ - \frac{\pi }{2},\frac{\pi }{2}]$, $1 \le i \le L_{\rm{r}}$.
Furthermore, we establish a path-response matrix (PRM) denoted as ${\bf{\Xi }} \in {{\mathbb {C}}^{L_r \times L_t}}$, which characterizes the response from the transmit reference position ${{\bf{t}}^0} = [0,0]$ to the receive reference position ${{\bf{r}}^0} = [0,0]$. Precisely, the entry in the $i$th row and $j$th column of ${\bf{\Xi }}$, denoted as $\alpha_{i,j}$, is the response coefficient between the $j$th transmit path and the $i$th receive path, where ${\alpha _{i,j}} \sim {\cal {CN}}(0,c/L)$, where $c=c_0d^{-\varpi}$ is the expected channel gain, $c_0$ is the unit distance path loss, $d$ is the distance between the transmit region and the receive region, and $\varpi$ is the path loss exponent. Consequently, the channel linking two antennas situated at ${\bf{t}}^0$ and ${\bf{r}}^0$ is expressed as
\begin{equation}
h({{\bf{t}}^0},{{\bf{r}}^0}) = {\bf{1}}_{{L_r}}^H{\bf{\Xi }}{{\bf{1}}_{{L_t}}},
\end{equation}
which represents the linear combination of all the components within the PRM.

Based on the fundamental principles of geometry depicted in Fig. \ref{Phase_system}, the phase difference between position ${\bf{r}}_{p,q}=[x_{{\rm{r}},p},y_{{\rm{r}},q}]$ and the reference point ${\bf{r}}^0$ is given by
\begin{equation}
\begin{aligned}
{\rho _{{\rm{r}},i}}({{\bf{r}}_{p,q}}) &= {e^{j\frac{{2\pi }}{\lambda }({x_{{\rm{r}},p}}\cos {\theta _{{\rm{r}},i}}\sin {\phi _{{\rm{r}},i}} + {y_{{\rm{r}},q}}\sin {\theta _{{\rm{r}},i}})}}\\
& = e^{j{{\bf{r}}_{p,q}}{{\bf{k}}_{\rm{r}}}({\theta _{{\rm{r}},i}},{\phi _{{\rm{r}},i}})}
\end{aligned},
\end{equation}
where ${{\bf{k}}_{\rm{r}}}({\theta _{{\rm{r}},i}},{\phi _{{\rm{r}},i}}) = \frac{{2\pi }}{\lambda }{[\cos {\theta _{{\rm{r}},i}}\sin {\phi _{{\rm{r}},i}},\sin {\theta _{{\rm{r}},i}}]^T}$, $1\le i \le L_r$ and $\lambda$ is the wavelength. Then, we define the field-response vector (FRV) in the receive region as
\begin{equation}
{\bf{f}}({\bf{r}}_{p,q}) = {[{{\rho _{{\rm{r}},1}}({\bf{r}}_{p,q})}},...,{{\rho _{{\rm{r}},i}}({\bf{r}}_{p,q})},...,{{\rho _{{\rm{r}},L_r}}({\bf{r}}_{p,q})}]^T.
\end{equation}

Similarly, for any position ${\bf{t}}_{m,n}=[x_{{\rm{t}},m},y_{{\rm{t}},n}]$ in the transmit region, the FRV is expressed as
\begin{equation}
{\bf{g}}({\bf{t}}_{m,n}) = {[{{\rho _{{\rm{t}},j}}({\bf{t}}_{m,n})}},...,{{\rho _{{\rm{t}},j}}({\bf{t}}_{m,n})},...,{{\rho _{{\rm{t}},L_t}}({\bf{t}}_{m,n})}]^T,
\end{equation}
where ${\rho _{{\rm{t}},j}}({{\bf{t}}_{m,n}}) = {e^{j{{\bf{k}}_{\rm{t}}}({\theta _{{\rm{t}},j}},{\phi _{{\rm{t}},j}}){{\bf{t}}_{m,n}}}}$, $1\le j \le L_t$. Therefore, the channel vector ${\bf h}({\bf{t}}_{m,n})\in {{\mathbb{C}}^{{N_{\rm{r}}} \times {1}}}$ between the FA antenna located at position ${\bf{t}}_{m,n}=[x_{{\rm{t}},m},y_{{\rm{t}},n}]$ and the fixed receive UPA can be obtained as
\begin{equation}\label{channel_model}
{\bf h}({\bf{t}}_{m,n}) = {\bf{F}}^H{\bf{\Xi g}}({\bf{t}}_{m,n}),
\end{equation}
where ${\bf{F}}=[{\bf{f}}({{\bf{r}}_{1,1}}),{\bf{f}}({{\bf{r}}_{1,2}}),...,{\bf{f}}({{\bf{r}}_{{N_1},{N_2}}})] \in {{\mathbb{C}}^{{L_r} \times {N_r}}}$ is the filed-response matrix (FRM) at the Rx side, which is a constant matrix since the receiver is equipped with a fixed UPA.

By combining channel vectors from all the $P$  FA positions to the fixed receive UPA, we can obtain the equivalent FA-IM channel matrix as
\begin{equation}
{\bf{H}}=[{\bf h}({\bf{t}}_{1,1}),...,{\bf h}({\bf{t}}_{m,n}),...,{\bf h}({\bf{t}}_{P_1,P_2})]\in{{\mathbb{C}}^{N_r \times P}}.
\end{equation}

\subsection{Signal model}
In the proposed scheme, the concept of IM is employed on the Tx side, where FA region grids are indexed instead of indexing FAs, i.e., $N_t$ out of $P$ grids are occupied by the $N_t$ transmit FAs to transmit signals. Note that all the $N_t$ FAs are activated simultaneously to achieve the maximum transmission rate for the given parameters $(P, N_t)$. As shown in Fig. \ref{System_Model}, the incoming bit stream is split into two sub-vectors of ${N_t}{\log _2}M$ and ${\lfloor{{\log }_2}{{{\tbinom{P}{N_t}}}}\rfloor}$ bits, denoted by ${\bf{b}}_1$ and  ${\bf{b}}_2$, respectively.  The modulated symbol bits in ${\bf{b}}_1$ are modulated to $N_t$ $M$-QAM symbols as ${\bf{s}}=[{s_1},...,{s_k},...,{s_{N_t}}]^T$, while the index bits in ${\bf{b}}_2$ are used to select a unique FA position pattern ${{\cal{I}}}=\{I_1,...,I_k,...,I_{N_t}\}\in {\mathbb{I}}$,  $I_k=\{1,2,...,P\}$, $k=\{1,2,...,N_t\}$ and $\mathbb{I}$ is the FA position pattern codebook, which will be carefully designed in Section II-C.
Thus, the equivalent transmit signal vector ${\bf{x}}\in {\mathbb{C}}^{P \times 1}$ is generated as
\begin{equation}
{\bf{x}} = {[\underbrace {0,...,0}_{{I_1} - 1},{s_1},\underbrace {0,...,0}_{{I_k} - {I_1} - 1},{s_k},\underbrace {0,...,0}_{{I_{{N_t}}} - {I_k} - 1},{s_{{N_t}}},\underbrace {0,...,0}_{P - {I_{{N_t}}}}]^T}.
\end{equation}

It is worth noting that a selected FA position pattern ${\cal{I}}=\{I_1,...,I_k,...,I_{N_t}\}$ corresponds to a unique FA position coordinate codebook ${{{\cal B}_{\cal I}}}=\{\Omega_{I_1},...,\Omega_{I_k},...,\Omega_{I_{N_t}}\}$. The FA position pattern to position coordinate mapping process can be expressed as
\begin{equation}\label{Map1}
f:I_k \to \Omega_{I_k}=(m,n) , \quad {\rm{i}}{\rm{.e}}{\rm{.,}}\quad {I_k} = (m - 1){P_2} + n,
\end{equation}
where $(m,n)$ is the Cartesian coordinate index of the position $I_k$ with $m=1,...,P_1$ and $n=1,...,P_2$.

To illustrate the mapping relationship more intuitively, a look-up table where $P=4$, $P_1=P_2=2$, and $N_t=2$ is described in Table I, which provides the corresponding FA position pattern $\cal I$ and FA position coordinate ${{{\cal B}_{\cal I}}}$ for the incoming index bits in ${\bf b}_2$. For example, when the incoming bits is 01, the corresponding FA position pattern is $\{1,3\}$, which means grid positions 1 and 3 are occupied by the $N_t=2$ transmit FAs. According to the mapping rule shown in \eqref{Map1}, the corresponding position coordinate can be obtained as $\{(1,1),(2,1)\}$. Note that this look-up table is random FA position pattern codebook, rather than optimal FA position pattern codebook.

{\renewcommand\arraystretch{1.5}
\begin{table}
\centering
{\caption{An example look-up table for $P=4$, $P_1=2$, $P_2=2$ and $N_t=2$.}}
\begin{tabular}{|c|c|c|}\hline
Bits in ${\bf{b}}_2$ & \thead{FA  position pattern \\ ${\cal{I}}=\{I_1,I_2\}$}  & \thead{FA position coordinate \\ ${{{\cal B}_{\cal I}}}=\{{\Omega_{I_1}},\Omega_{I_2}\}$}    \\\hline\hline
00 & $\{1,2\}$ &  $\{(1,1),(1,2)\}$  \\\hline
01 & $\{1,3\}$ &  $\{(1,1),(2,1)\}$   \\\hline
10 & $\{2,4\}$ &  $\{(1,2),(2,2)\}$   \\\hline
11 & $\{3,4\}$ &  $\{(2,1),(2,2)\}$   \\\hline
unused & $\{1,4\}$ &  $\{(1,1),(2,2)\}$   \\\hline
unused & $\{2,3\}$ &  $\{(1,2),(2,1)\}$  \\\hline
\end{tabular}
\end{table}}

After obtaining the corresponding FA position coordinate codebook ${{{\cal B}_{\cal I}}}$, the effective channel ${\bf{H}}_{\rm{eff}}^{{{\cal B}_{\cal I}}}\in {\mathbb{C}^{N_r \times N_t}}$ can be represented as
\begin{equation}\label{receive_1}
{\bf{H}}_{\rm{eff}}^{{{\cal B}_{\cal I}}}=[{\bf h}({\bf{t}}_{\Omega_{I_1}}),...,{\bf h}({\bf{t}}_{\Omega_{I_k}}),...,{\bf h}({\bf{t}}_{\Omega_{I_{N_t}}})].
\end{equation}
Thus, the received signal ${\bf{y}} \in {\mathbb{C}}^{N_r \times 1}$ can be written as
\begin{equation}
{\bf{y}} ={\bf{Hx}}+ {\bf{n}}=\sum\limits_{k = 1}^{{N_t}} {{\bf{h}}({{\bf{t}}_{{\Omega _{{I_k}}}}}){s_k}} + {\bf{n}},
\end{equation}
where ${\bf{n}}\in {\mathbb{C}}^{N_r \times 1}\sim {\cal{CN}}(0,{N_0}{\bf{I}}_{N_r})$ is the additive white Gaussian noise. In this paper, we assume that the perfect channel state information (CSI) is available at receiver, and the effect of imperfect CSI will be discussed in Section V{\footnote{Channel estimation for the FA-assisted MIMO systems is discussed in \cite{ma2023compressed}, where compressed sensing based method is employed to reconstruct the channel by exploiting the sparse representation of the field-response based channel model.}}.

\subsection{FA position pattern codebook design}
According to the index bits mapping rule, it is obvious that there are totally $A={{{\tbinom{P}{N_t}}}}$ possible FA position patterns available. However, the number of FA position patterns we need to use here is $K=2^{\lfloor \log_2 A \rfloor}$ due to the fact that the indices of FA position patterns are mapped to binary-bit blocks, thus these FA position patterns can transmit ${\lfloor \log_2 A \rfloor}$ bits per channel use (bpcu). For example, we can see from Table I that there are $A={{{\tbinom{4}{2}}}}=6$ possible FA position patterns, but only $K=2^{\lfloor \log_2 6 \rfloor}=4$ can be used. Thus, it is necessary to select $K$ out of $A$ FA position patterns for indexing to further improve system performance.

Let's define the FA position pattern set as ${{\cal L}=\{{\cal{I}}_1,...,{\cal{I}}_i,...,{\cal{I}}_A\}}$, $i=\{1,2,...,A\}$, where the effective channel of the $i$th FA position pattern is given as
\begin{equation}
 {\bf{H}}_{\rm{eff}}^{{\cal I}_i}={\bf{H}}_{\rm{eff}}^{{{\cal B}_{{\cal I}_i}}}=[{\bf h}({\bf{t}}_{\Omega_{I_{i,1}}}),...,{\bf h}({\bf{t}}_{\Omega_{I_{i,k}}}),...,{\bf h}({\bf{t}}_{\Omega_{I_{i,N_t}}})].   
\end{equation}

For the given effective channel realization and SNR, the capacity of the proposed FA-IM system is bounded as \cite{rajashekar2013antenna}
\begin{equation}
\beta  \le {{\cal C}_{{\rm{FA - IM}}}} \le \beta  + {\log _2}({N_t}),
\end{equation}
where $\beta  = \frac{1}{{{N_t}}}{\log _2}(1 + \frac{{{{\left\| {\bf H_{{\rm{eff}}}^{\cal I}} \right\|}^2}}}{{{N_0}}})$. It is clear that $\beta$ is maximized by choosing $N_t$ positions corresponding to the largest channel norms out of the $P$ positions{\footnote{{Maximizing the Euclidean distance among indexed FA position patterns can help the system to further boost performance, but this algorithm suffers from high computational complexity.}}}. Hence, the FA position pattern codebook can be designed as 
\begin{equation}
{\mathbb{I}}=\{{\cal I}_1,{\cal I}_2,...,{\cal I}_K\}, 
\end{equation}
which satisfies that
\begin{equation}
{\left\| {{\bf{H}}_{{\rm{eff}}}^{{{\cal I}_1}}} \right\|^2} > {\left\| {{\bf{H}}_{{\rm{eff}}}^{{{\cal I}_2}}} \right\|^2} > ... > {\left\| {{\bf{H}}_{{\rm{eff}}}^{{{\cal I}_K}}} \right\|^2} > ... > {\left\| {{\bf{H}}_{{\rm{eff}}}^{{{\cal I}_A}}} \right\|^2}.
\end{equation}

\section{Proposed Low-complexity Detector}
The optimal ML detector is adopted to jointly detect the FA position pattern as well as the modulated symbols by exhaustively searching all the possible transmitted signal vectors. The output of the ML detector is given by
\begin{equation}\label{ML_detector}
[\hat {\cal I},{\bf{\hat s}}] = \arg \mathop {\min }\limits_{{\cal I},{\bf{s}}} {\left\| {{\bf{y}} - \sum\limits_{k = 1}^{{N_t}} {{\bf{h}}({{\bf{t}}_{{\Omega _{{I_k}}}}}){s_k}} } \right\|^2}.
\end{equation}
Note that the ML detector needs to evaluate all the FA position patterns together with the symbols, i.e., the complexity order of the ML detector is ${\cal{O}}(KM^{N_t})$, leading to high computational complexity as the value of $P$, $N_t$ and $M$ increase. 

\subsection{Proposed efficient sparse Bayesian detector}
In this subsection, we propose a low-complexity compress sensing based detector, called efficient sparse Bayesian detector, which advances the original sparse Bayesian algorithm \cite{ji2008bayesian} by simplifying the iteration process via taking the inherent sparsity of the proposed FA-IM scheme into consideration.

The objective of the proposed efficient sparse Bayesian detector is to identify the selected FA position pattern, i.e., the non-zero locations of the transmit signal vector $\bf{x}$. We embark upon our discourse by introducing a prior concerning the vector $\bf x$, the probabilistic characterization of which is governed by an array of hyperparameters denoted by the vector ${\boldsymbol{\gamma }} = {[{\gamma _1},...,{\gamma _P}]^T}$. This discernibly implies that  the Gaussian prior $\gamma_i$ is specifically assigned to the $i$th element of $\bf x$, where the Gaussian prior is $p({\bf{x}}\left| {\boldsymbol{\gamma }}  \right.) = {\cal{CN}}({\bf{0}},{\bf{\Gamma }})$ with ${\bf{\Gamma}}={\rm{diag}}({\boldsymbol{\gamma }})$ \cite{al2017gamp}. Then, the estimated ${\hat {\boldsymbol{\gamma }}}$ can be obtained by solving the following maximum a posterior (MAP) problem
\begin{equation}\label{ESB_1}
{\hat {\boldsymbol{\gamma }} } = \arg \mathop {\min }\limits_{\boldsymbol{\gamma }} p({{\boldsymbol{\gamma }}\left| {{\bf{y}}} \right.}).
\end{equation}
According to the Bayesian rule, \eqref{ESB_1} can be further rewritten as
\begin{equation}\label{ESB_2}
{\hat {\boldsymbol{\gamma }}} = \arg \mathop {\min }\limits_{{\boldsymbol{\gamma }}} p( {{\bf{y}}\left| {\boldsymbol{\gamma }} \right.})p({\boldsymbol{\gamma }}),
\end{equation}
where $p({ {\bf{y}}\left|{\boldsymbol{\gamma }} \right.})$ is the likelihood function of ${{\bf{y}}}$ given ${\boldsymbol{\gamma }}$. It is calculated that $p({ {\bf{y}}\left|{\boldsymbol{\gamma }} \right.})$ satisfies Gaussian distributions ${\cal{CN}}({\bf{0}},{\bf{\Sigma_{\bf y}}})$, where ${\bf{\Sigma_{\bf y}}}=N_0{\bf{I}}_{N_r}+{\bf {H \Gamma  H}}^H$. By combining the likelihood function and the prior, the posterior density of $\bf x$ can be written as $p({\bf{x}}\left| {\bf{y}} \right.,{\boldsymbol{\gamma }})={\cal{N}}({\boldsymbol{\mu }},{\bf{\Sigma }})$ with mean and covariance given by
\begin{equation}\label{ESB_3}
{\boldsymbol{\mu }} = {\bf{\Gamma }}{{\bf{H}}^H}{{\bf{\Sigma }}_{\bf{y}}}{\bf{y}}
\end{equation}
and
\begin{equation}\label{ESB_4}
{\bf{\Sigma }} = {\bf{\Gamma }} - {\bf{\Gamma }}{{\bf{H}}^H}{\bf{\Sigma }}_{\bf{y}}^{ - 1}{\bf{H\Gamma }},
\end{equation}
respectively. Once the hyperparameter vector is determined, the transmit signal vector $\bf x$ can be estimated by
\begin{equation}
{\bf{\hat x}} = \arg \mathop {\max }\limits_{\bf{x}} p({\bf{x}}\left| {\bf{y}} \right.,{\boldsymbol{\gamma }}).
\end{equation}
To facilitate the computation of ${\boldsymbol{\mu }}$ and ${\bf{\Sigma }}$, the determination of the hyperparameter ${\boldsymbol{\gamma }}$ is imperative as shown in \eqref{ESB_3} and \eqref{ESB_4}. As the estimation of ${\boldsymbol{\gamma }}$ also depends on ${\boldsymbol{\mu }}$ and ${\bf{\Sigma }}$, an iterative approach is needed. The estimation of ${\boldsymbol{\gamma }}$ is a typical type-II ML problem and we employ the update rule in \cite{wipf2007empirical}, given by
\begin{equation}\label{ESB_5}
\gamma _i^{({\rm{new}})} = {\left\| {{\mu _i}} \right\|^2} + {\Sigma _{i,i}},\quad i = 1,...,P,
\end{equation}
where $\gamma _i$ is the $i$th entry of ${\boldsymbol{\gamma }}$, ${\mu _i}$ and ${\Sigma _{i,i}}$ refer to the $i$th entry of ${\boldsymbol{\mu }}$ and the $i$th diagonal entry of $\bf{\Sigma}$, respectively. The update of \eqref{ESB_3}, \eqref{ESB_4} and \eqref{ESB_5} is repeated until the convergence, and then an estimate of ${\boldsymbol{\gamma }}$ is obtained.

In original sparse Bayesian algorithm, the elements in  ${\boldsymbol{\gamma }}$ undergo a process of sequential updates at each iteration. However, since the hyperparameter vector ${\boldsymbol{\gamma }}$ shares the same spare profile as $\bf x$, i.e., the majority of elements in ${\boldsymbol{\gamma }}$ are zeros, the updating process of those zero positions can be safely left out. Under this circumstance, we modify the iteration process to reduce the  computational complexity by pruning the hyperparameter vector ${\boldsymbol{\gamma }}$, which will be detailed in the sequel.

As mentioned above, the $i$th element in hyperparameter vector ${\boldsymbol{\gamma }}$ indicates the probabilistic characterization of the corresponding $i$th element in $\bf x$. That is to say, the high value of $\gamma_i$ means that the $i$th position is more likely to be occupied by the FA, termed as active position. As a result, in the modified iteration process, ${\psi _t}$ positions characterized by lower probabilities are designated as inactive positions. These inactive positions are decisively discarded from the current FA position candidate set ${\tilde {\cal I}}$.  Therefore, the corresponding hyperparameter vector and channel matrix can be pruned as ${\boldsymbol{\gamma }}=[{\boldsymbol{\gamma }}]_{:,{\tilde {\cal I}}}$ and ${\bf{H}}=[\bf{H}]_{:,{\tilde {\cal I}}}$, respectively. Additionally, the scale of new hyperparameter vector is thus updated by
$P^*=P^*-{\psi _t}$. It is pertinent to emphasize that after $T_b$ iteration, the final FA position candidate set remains invariant, comprising only $P^*$ elements, which can be calculated by
\begin{equation}\label{Value_P}
{P^*} = P - \sum\limits_{t = 1}^{{T_b}} {{\psi _t}}.
\end{equation}
To avoid the exclusion of non-zero candidates, we set $P^* > P-N_t$, considering the inherent sparsity of the proposed FA-IM system.

The procedure of the proposed efficient sparse Bayesian detector is summarized in \textbf{Algorithm 1}. After $T_{\rm{max}}$ iterations, we can obtain the estimated FA position pattern by $\hat {\cal I}={\bf{L}}(1:N_t)$, where ${\bf{L}}={\rm{sort}}_{\rm{descent}}({\tilde {\cal I}})$ and ${\rm{sort}}_{\rm{descent}}(\bf{a})$ is to sort the elements of $\bf a$ in a descending order and returns the positions. Finally, the bit information is detected by FA-IM demodulation.

\begin{algorithm}[t]
\caption{Proposed efficient sparse Bayesian detector for the FA-IM system}
\begin{algorithmic}[1]
\REQUIRE ~~\\
The received signal: ${\bf{y}}$;\\
Equivalent channel matrix: ${\bf{H}}$;\\
Number of FA and positions: $N_t$ and $P$;\\
The maximum number of iterations: $T_{\rm max}$;\\
The parameters for pruning the hyperparameter vector: \\
$\Psi  = \left\{ {{\psi _1},...,{\psi _t},...,{\psi _{{T_b}}}} \right\}$.\\
\ENSURE $\hat {\cal I}$, ${\bf{\hat s}}$
\STATE (Iteration initialization): The iterative index $t=1$, the initial hyperparameter vector ${\boldsymbol{\gamma }}={\bf{1}}_P$, the initial FA position candidate ${\tilde {\cal I}}=\{1,2,..,P\}$, the initial scale of hyperparameter vector $P^{*}=P$.\\
\STATE \textbf{for} $t = 1,2,...,T_{\rm{max}}$, \textbf{do}
\STATE \quad Update ${\boldsymbol{\mu }}$ according to \eqref{ESB_3};
\STATE \quad Update ${\bf{\Gamma }}$ according to \eqref{ESB_4};
\STATE \quad Update ${\boldsymbol{\gamma }}$ according to \eqref{ESB_5};
\STATE \quad \textbf{while} $t \le T_b$, \textbf{do}
\STATE \qquad $P^{*}=P^{*}-{\psi _t}$;
\STATE \qquad ${\tilde {\cal I}}=\max (| {\boldsymbol{\gamma }}|,{P^*})$;
\STATE \qquad ${\bf{H}}=[\bf{H}]_{:,{\tilde {\cal I}}}$, ${\boldsymbol{\gamma }}=[{\boldsymbol{\gamma }}]_{:,{\tilde {\cal I}}}$;
\STATE \quad \textbf{end while}
\STATE \textbf{end for}
\STATE ${\bf{L}}={\rm{sort}}_{\rm{descent}}({\tilde {\cal I}})$;
\STATE ${\hat {\cal I}}={\bf{L}}(1:N_t)$; $\hat {\bf s}=({\bf{H}}_{:,\hat {\cal I}})^{\dag}{\bf y}$.
\end{algorithmic}
\end{algorithm}

{\renewcommand\arraystretch{1.5}
\begin{table}
\centering
{\caption{Complexity of common operations in terms of flops}}
{\begin{tabular}{|l|l|}\hline
Operations & Real-valued flops\\\hline
${\bf{AB}}$ & $2mp(4n-1)$  \\\hline
${\bf{C}}^{-1}$ & $\frac{1}{2}{n^3} + \frac{3}{2}n$  \\\hline
$\left\| {\bf{c}} \right\|^2$ & $2n$  \\\hline
${\bf{c}}\pm{\bf{d}}$ & $4n-1$  \\\hline
\end{tabular}}
\end{table}}

\subsection{Complexity analysis}
In this subsection, the computational complexity is quantified in terms of the real-valued flops, encompassing real-valued multiplications and additions. For the specific matrices ${\bf{A}} \in {{\mathbb{C}}^{m \times n}}$, ${\bf{B}} \in {{\mathbb{C}}^{n \times p}}$, ${\bf{C}} \in {{\mathbb{C}}^{n \times n}}$, ${\bf{c}} \in {{\mathbb{C}}^{n \times 1}}$ and ${\bf{d}} \in {{\mathbb{C}}^{n \times 1}}$,  Table II presents the flops required for the operations involved in the detection algorithm.

Accordingly, the detection complexity of the optimal ML detector (i.e., \eqref{ML_detector}) is expressed as
\begin{equation}
C_{\rm{ML}} =KM^{N_t}(12N_rN_t+2N_r),    
\end{equation}
where $KM^{N_t}$ is the total search space of the proposed FA-IM scheme, and $12N_rN_t+2N_r$ is the complexity of calculating ${\| {{\bf{y}} - \sum\limits_{k = 1}^{{N_t}} {{\bf{h}}({{\bf{t}}_{{\Omega _{{I_k}}}}}){s_k}} } \|^2}$ in \eqref{ML_detector}.

The complexity of the proposed efficient sparse Bayesian detector is calculated by
\begin{equation}
\begin{aligned}
{C_{{\rm{Proposed}}}}& = {T_{\max }}[\underbrace {16N_r^2{P^ * } + 12{N_r}{P^ * } - 2{P^ * }}_{{\rm{Update}} \;{\boldsymbol{\mu }}  } + \underbrace {2{P^ * }}_{{\rm{Update}}\;{\boldsymbol{\gamma }} }\\
&+\underbrace {8N_r^2{P^ * } + 4{{({P^ * })}^2}(2{N_r} - 1) + \frac{1}{2}N_r^3 + \frac{3}{2}{N_r}}_{{\rm{Update}}\;{\bf{\Gamma }} }]\\
&={T_{\max }}[24N_r^2{P^ * }+4{{({P^ * })}^2}(2{N_r} - 1)+12{N_r}{P^ * } \\
&\qquad \qquad + \frac{1}{2}N_r^3 + \frac{3}{2}{N_r}],
\end{aligned}
\end{equation}
where ${P^ * }$ is the scale of hyperparameter vector ${\boldsymbol{\gamma }} $, which will be updated according to \eqref{Value_P}.

Table III presents the computational complexity of the proposed detector under different parameter settings. It is apparent that all the sparse Bayesian detector achieve significant complexity reduction (i.e., 99.9$\%$) compared to the optimal ML detector. Furthermore, the proposed efficient sparse Bayesian algorithm saves nearly 26$\%$ and 25$\%$ when the parameter settings $(N_t,N_r,P,M)$ are $(4,20,20,4)$ and $(8,32,32,4)$  compared to the original sparse Bayesian algorithm \cite{wipf2007empirical}, respectively. This complexity reduction benefits from the proposed detector simplifying the iteration process by pruning the hyperparameter vector. More importantly, thanks to the candidate pruning mechanism, many inactive elements can obtain their own optimal solutions (i.e., 0) in advance, eliminating the need to run the iterative algorithm until convergence. It contributes to the enhanced detection performance by reducing the interference, as will be evidenced by our simulation results\footnote{The computational complexity of the proposed detector can be further reduced by replacing the matrix inversion operations in \eqref{ESB_4} during the iteration process. For example, the Neumann series method can be employed to approximate the large-scale matrix inversion \cite{zhang2022low}, and it will be left for our future work.}.

{\renewcommand\arraystretch{1.5}
\begin{table}
\centering
{\caption{The computational complexity comparison of different detectors for the FA-IM scheme}}
\begin{tabular}{|l|l|l|}\hline
\multirow{2}{*}{Detectors}&\multicolumn{2}{c|}{$(N_t,N_r,P,M)$}\\
\cline{2-3}
 & (4,20,20,4)&  (8,32,32,4)\\ \hline
Optimal ML& $1.1 \times 10^{9}$ & $1.7\times 10^{15}$ \\ \hline
Original sparse Bayesian \cite{wipf2007empirical}  & $2.6\times 10^{6}$  &$1.6 \times 10^{7}$ \\ \hline
Proposed efficient sparse Bayesian & $1.9 \times 10^{6}$ &$1.2 \times 10^{7}$  \\ \hline
\end{tabular}
\end{table}}

\section{Performance Analysis}
\subsection{Spectral efficiency analysis}
This subsection analyzes the bits per channel use (bpcu) throughput of the proposed FA-IM, where the VBLAST based FAs-assisted MIMO without IM (namely FA-VBLAST) and the classic FPAs-assisted MIMO with IM (namely SM-MIMO) serve as its benchmarks to demonstrate the superiority of the proposed scheme.

The SE of the classic SM-MIMO and the FA-VBLAST are given by
\begin{equation}
R_{\rm{FA-VBLAST}}=N_t{\log _2}M   
\end{equation}
and
\begin{equation}
R_{\rm{SM-MIMO}}={\log _2}M+{{\log }_2}{N_t},
\end{equation}
respectively.

\begin{figure}[t]
\centering
\includegraphics[width=0.4\textwidth]{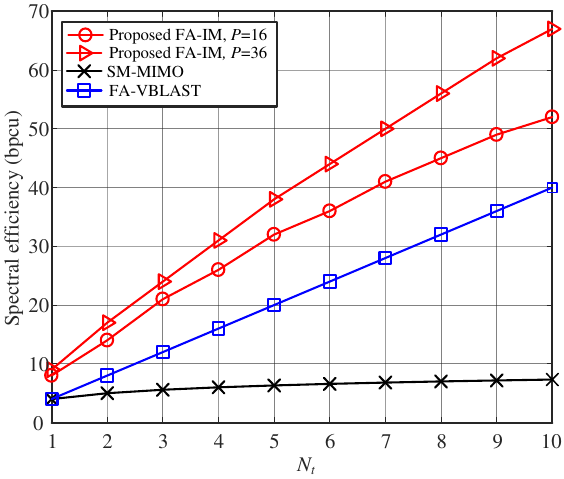}
\caption{Spectral efficiency comparisons between the proposed FA-IM with the conventional FA-VBLAST and SM-MIMO systems.}
\label{SE}
\end{figure}

Based on the signal model described in Section II-B, the SE of the proposed FA-IM is calculated by
\begin{equation}\label{data_rate_1}
R_{\rm{FA-IM}}={N_t}{\log _2}M+{\lfloor{{\log }_2}{{{\tbinom{P}{N_t}}}}\rfloor}.
\end{equation}
To illustrate the SE performance more intuitively, Fig. \ref{SE} plots the SE of the proposed FA-IM and its benchmarks for various number of transmit antennas. The results clearly show that the proposed scheme achieves higher SE compared to its benchmark systems when equipped with the same number of transmit antennas. More precise, the extent of SE improvement is contingent upon the number of the positions $P$. In other words, as the value of  $P$ increases, the FA position patterns conveys more information bits,  thus offering a significant SE enhancement without increasing any hardware cost.

\subsection{ABER performance analysis}
In this subsection, we derive a closed-form expression for the upper bound on ABEP to validate the proposed system. According to the well-known union bounding technique \cite{simon2001digital}, the ABEP of the proposed FA-IM system is bounded as
\begin{equation}
{\rm{ABEP}} \le \frac{1}{{R{2^R}}}\sum\limits_{{\cal I},{\bf{s}}} {\sum\limits_{\hat {\cal I},{\bf{\hat s}}} d\{({\cal I},{\bf{s}}) \to (\hat {\cal I},{\bf{\hat s}})\} } {\mathbb{P}}\{({\cal I},{\bf{s}}) \to (\hat {\cal I},{\bf{\hat s}})\},
\end{equation}
where $R$ is the date rate given in \eqref{data_rate_1},  $d\{({\cal I},{\bf{s}}) \to (\hat {\cal I},{\bf{\hat s}})\}$ is the total number of erroneous bits and ${\mathbb{P}}\{({\cal I},{\bf{s}}) \to (\hat {\cal I},{\bf{\hat s}})\}$ is the unconditional pairwise error probability (UPEP).

To obtain the upper bound, we first compute the conditional PEP (CPEP) as
\begin{equation}\label{PEP_1}
\begin{aligned}
&{\mathbb{P}}\{ ({\cal{I}},{\bf{s}}) \to (\hat{\cal{I}},{\bf{\hat s}})\left| {{\alpha _{1,1}},...,{\alpha _{L_r,L_t}} }\right\} \\
& \quad=  {\mathbb{P}}\left\{ {\left\| {{\bf{y}} - \sum\limits_{k = 1}^{{N_t}} {{\bf{h}}({{\bf{t}}_{{\Omega _{{I_k}}}}}){s_k}} } \right\|^2} > {\left\| {{\bf{y}} - \sum\limits_{k = 1}^{{N_t}} {{\bf{h}}({{\bf{t}}_{{\Omega _{{{\hat I}_k}}}}}){{\hat s}_k}} } \right\|^2}\right\}\\
&\quad={\mathbb{P}}\left\{ \zeta < 0\right\},
\end{aligned}
\end{equation}
where 
\begin{equation}\label{PEP_5}
\begin{aligned}
\zeta & = {\left\| {\sum\limits_{k = 1}^{{N_t}} {\left( {{\bf{h}}({{\bf{t}}_{{\Omega _{{I_k}}}}}){s_k} - {\bf{h}}({{\bf{t}}_{{\Omega _{{{\hat I}_k}}}}}){{\hat s}_k}} \right)} } \right\|^2} + \\
&\quad 2\Re \left\{ {{{\bf{n}}^H}\left( {\sum\limits_{k = 1}^{{N_t}} {\left( {{\bf{h}}({{\bf{t}}_{{\Omega _{{I_k}}}}}){s_k} - {\bf{h}}({{\bf{t}}_{{\Omega _{{{\hat I}_k}}}}}){{\hat s}_k}} \right)} } \right)} \right\}.
\end{aligned}   
\end{equation}
Let $\Delta=\sum\limits_{k = 1}^{{N_t}} {\left( {{\bf{h}}({{\bf{t}}_{{\Omega _{{I_k}}}}}){s_k} - {\bf{h}}({{\bf{t}}_{{\Omega _{{{\hat I}_k}}}}}){{\hat s}_k}} \right)}$ for ease of notation, thus $\zeta$ can be rewritten as $\zeta=\|\Delta\|^2+2\Re\{ {{\bf{n}}^H}\Delta\}$,
which is a Gaussian random variable with mean ${\mu _\zeta} $ and variance $\sigma _\zeta ^2$, i.e., $\zeta  \sim {\cal{N}}({\mu _\zeta },\sigma _\zeta ^2)$. The mean ${\mu _\zeta }$ can be computed directly as
${\mu _\zeta } = {\mathbb{E}}(\|\Delta\|^2+2\Re\{ {{\bf{n}}^H}\Delta\})= {\| \Delta\|^2}$ due to ${\mathbb{E}}(2\Re\{ {{\bf{n}}^H}\Delta\})=0$. On the other hand, the variance can be manipulated as
\begin{equation}\label{APP_1}
\begin{aligned}
\sigma _\zeta ^2& = {\mathbb{E}}[{(\zeta  - {\mu _\zeta })^2}]\\
& = {\mathbb{E}}[(2\Re\{ {{\bf{n}}^H}\Delta\})^2]\\
& = {\mathbb{E}}(4\Re \{ {({{\bf{n}}^H})^2}\} ){\mathbb{E}}(\Re \{\| \Delta\|^2\})\\
& = 2{N_0}\| \Delta\|^2.
\end{aligned}
\end{equation}
Therefore, the CPEP can be further calculated as
\begin{equation}
\begin{aligned}
&{\mathbb{P}}\{ ({\cal{I}},{\bf{s}}) \to (\hat {\cal{I}},{\bf{\hat s}})\left| {{\alpha _{1,1}},...,{\alpha _{L_r,L_t}} } \right.\}\\
&\qquad =Q\left( {\sqrt {\frac{\| \Delta\|^2}{{2{N_0}}}} } \right)=Q\left( {\sqrt {\frac{{Z}}{{2{N_0}}}} } \right)
\end{aligned}
\end{equation}
where $Z=\| \Delta\|^2$ and $Q(x)$ is the Gaussian $Q$-function.

By approximating the $Q$-function as $Q(x) \approx   \frac{1}{{12}}\exp ( - {{{x^2}}}/{2}) + \frac{1}{4}\exp ( - {{2{x^2}}}/{3})$ \cite{ma2020large}, the CPEP can be approximated as
\begin{equation}
\begin{aligned}
&{\mathbb{P}}\{ ({\cal{I}},{\bf{s}}) \to (\hat {\cal{I}},{\bf{\hat s}})\left| {{\alpha _{1,1}},...,{\alpha _{L_r,L_t}} } \right.\}   \\
&\qquad \qquad \approx  \frac{1}{{12}}\exp \left( - \frac{Z}{{4{N_0}}}\right) + \frac{1}{4}\exp \left( - \frac{Z}{{3{N_0}}}\right).
\end{aligned}
\end{equation}
Furthermore, the UPEP can be obtained as
\begin{equation}\label{UPEP_3}
\begin{aligned}
&{\mathbb{P}}\{ ({\cal{I}},{\bf{s}}) \to (\hat {\cal{I}},{\bf{\hat s}})\} \\
&\quad \approx {{\mathbb{E}}_Z}\left[ {\frac{1}{{12}}\exp \left( { - \frac{Z}{{4{N_0}}}} \right)} \right] + {{\mathbb{E}}_Z}\left[ {\frac{1}{4}\exp \left( { - \frac{Z}{{3{N_0}}}} \right)} \right]\\
&\quad= \frac{1}{{12}}{M_Z}\left( { - \frac{1}{{4{N_0}}}} \right) + \frac{1}{4}{M_Z}\left( { - \frac{1}{{3{N_0}}}} \right),
\end{aligned}
\end{equation}
where $M_Z(\varepsilon)={\mathbb{E}}_Z[\exp(\varepsilon Z)]$ is the moment-generating function (MGF) of $Z$. 

Without loss of generality, we employ geometry channel model in this work, where the number of transmit and receive paths are the same, i.e., $L_t=L_r=L$. Next, we derive the UPEP expression in two cases, i.e., finite-path case and infinite-path case.

\subsubsection{Finite-path channel case}
In the finite-path channel case, it is difficult to derive the explicit expression of UPEP, we can obtain the approximate UPEP by the Monte Carlo method.

\textbf{Lemma 1.} \textit{In this case of finite-path channel, the analytical expression for UPEP can be expressed as}
\begin{equation}\label{UPEP_1}
\begin{aligned}
&{\mathbb{P}}\{ ({\cal{I}},{\bf{s}}) \to (\hat {\cal{I}},{\bf{\hat s}})\}\\
&\approx  {\mathbb{E}}\left\{\frac{1}{{12}}{\left( {1 + \frac{L}{{2{N_0}B}}} \right)^{ - L}} + \frac{1}{4}{\left( {1 + \frac{2L}{{3{N_0}B}}} \right)^{ -L}}\right\},
\end{aligned}
\end{equation}
where 
\begin{equation}
B = \left\{ {\begin{array}{*{20}{l}}
{{{\left| {{s_k} - {{\hat s}_k}} \right|}^2}{{\left| {\sum\limits_{k = 1}^{{N_t}} {{{\bf{F}}^H}{\bf{g}}({{\bf{t}}_{{\Omega _{{{\cal I}_k}}}}})} } \right|}^2},}&{{\cal I} = \hat {\cal I}}\\
{\left| {{{\bf{F}}^H}\sum\limits_{k = 1}^{{N_t}} {\left( {({\bf{g}}({{\bf{t}}_{{\Omega _{{{\cal I}_k}}}}}){s_k} - {\bf{g}}({{\bf{t}}_{{\Omega _{{{\hat {\cal I}}_k}}}}}){{\hat s}_k})} \right)} } \right|,}&{{\cal I} \ne \hat {\cal I}}
\end{array}} \right..
\end{equation}
\textit{Proof:} See Appendix A.

\subsubsection{Infinite-path channel case, i.e., $L \to \infty$}
In the case of infinite-path, the channel model in \eqref{channel_model} is equal to correlated Rayleigh fading channel \cite{zhu2022modeling}.

\textbf{Lemma 2.} \textit{In this case of infinite-path channel, the analytical expression for UPEP can be expressed as}
\begin{equation}\label{UPEP_11}
\begin{aligned}
{\mathbb{P}}\{ ({\cal{I}},{\bf{s}}) \to (\hat {\cal{I}},{\bf{\hat s}})\} &\approx \frac{1}{{12}}{\left[ {\det \left( {{{\bf{I}}_{{N_r}}} + \frac{L}{{2{N_0}{\bf{R}}{{\left| {{s_k} - {{\hat s}_k}} \right|}^2}}}} \right)} \right]^{ - N_t}}\\
&\quad + \frac{1}{4}{\left[ {\det \left( {{{\bf{I}}_{{N_r}}} + \frac{{2L}}{{3{N_0}{\bf{R}}{{\left| {{s_k} - {{\hat s}_k}} \right|}^2}}}} \right)} \right]^{ - N_t}},
\end{aligned}
\end{equation}
where $\bf R$ is the spatial correlation matrix of channel $\bf H$.

\textit{Proof:} See Appendix B.

\textit{Remark 1 (Diversity gain of the system):} We first rewrite \eqref{UPEP_11} as
\begin{equation}\label{UPEP_D}
\begin{aligned}
{\mathbb{P}}\{ ({\cal{I}},{\bf{s}}) \to (\hat {\cal{I}},{\bf{\hat s}})\} \approx& \frac{1}{{12}}{\left[ {\det \left( {{{\bf{I}}_{{N_r}}} + {\rho _1}{\bf{B}}} \right)} \right]^{ - {N_t}}}\\
&+ \frac{1}{4}{\left[ {\det \left( {{{\bf{I}}_{{N_r}}} + {\rho _2}{\bf{B}}} \right)} \right]^{ - {N_t}}},
\end{aligned}
\end{equation}
where $\rho_1=\frac{1}{2{N_0}}$, $\rho_2=\frac{2}{3N_0}$ and ${\bf{B}} = \frac{L}{{{\bf{R}}{{\left| {{s_k} - {{\hat s}_k}} \right|}^2}}}$. Define ${\bf A}_i={{{\bf{I}}_{{N_r}}} + {\rho _i}{\bf{B}}}$ for $i=1,2$, we have
\begin{equation}
\det ({{\bf{A}}_i}) = \prod\limits_{\eta  = 1}^\Upsilon  {{\kappa _\eta }} ({{\bf{A}}_i}) = \prod\limits_{\tau  = 1}^r {(1 + {\rho _i}{\kappa _\tau }} ({\bf{B}})),
\end{equation}
where $r$ is the rank of ${\bf{B}}$ and ${\kappa _\tau } ({\bf{B}})$ is the $\tau$th non-zero eigenvalue of $\bf B$. For high SNR values, i.e., ${\rho _i} \gg 1$, \eqref{UPEP_D} can be further simplified as
\begin{equation}
\begin{aligned}
&{\mathbb{P}}\{ ({\cal{I}},{\bf{s}}) \to (\hat {\cal{I}},{\bf{\hat s}})\}  \\
&\approx \frac{1}{{12}}{\left(\rho _1^r\prod\limits_{\tau  = 1}^r {{\kappa _\tau }} ({\bf{B}})\right)^{ - {N_t}}} + \frac{1}{4}{\left(\rho _2^r\prod\limits_{\tau  = 1}^r {{\kappa _\tau }} ({\bf{B}})\right)^{ - {N_t}}}\\
&=\left( {\frac{1}{{12}}\rho _1^{ - r{N_t}} + \frac{1}{4}\rho _2^{ - r{N_t}}} \right){\left( {\prod\limits_{\tau  = 1}^r {{\kappa _\tau }} ({\bf{B}})} \right)^{ - {N_t}}}.
\end{aligned}
\end{equation}

In infinite-path channel case, the diversity gain $\cal D$ of the proposed FA-IM scheme is calculated as
\begin{equation}
\begin{aligned}
{\cal D} &= \mathop {\lim }\limits_{\rho  \to \infty }  - \frac{{{{\log }_2}\left( {{\mathbb{P}}\{ ({\cal{I}},{\bf{s}}) \to (\hat {\cal{I}},{\bf{\hat s}})\}} \right)}}{{{{\log }_2}\rho }}\\
&=\mathop {\lim }\limits_{\rho  \to \infty }  - \frac{{{{\log }_2}\left( {\left( {\frac{1}{{12}}\rho _1^{ - r{N_t}} + \frac{1}{4}\rho _2^{ - r{N_t}}} \right){{\left( {\prod\limits_{\tau  = 1}^r {{\kappa _\tau }} ({\bf{B}})} \right)}^{ - {N_t}}}} \right)}}{{{{\log }_2}\rho }}\\
&=\mathop {\lim }\limits_{\rho  \to \infty } \frac{{r{N_t}{{\log }_2}\rho  + {N_t}{{\log }_2}\left( {\prod\limits_{\tau  = 1}^r {{\kappa _\tau }} ({\bf{B}})} \right)}}{{{{\log }_2}\rho }}=rN_t,
\end{aligned}
\end{equation}
where $\rho=\frac{1}{N_0}$.

\begin{figure}[t]
\centering
\includegraphics[width=0.4\textwidth]{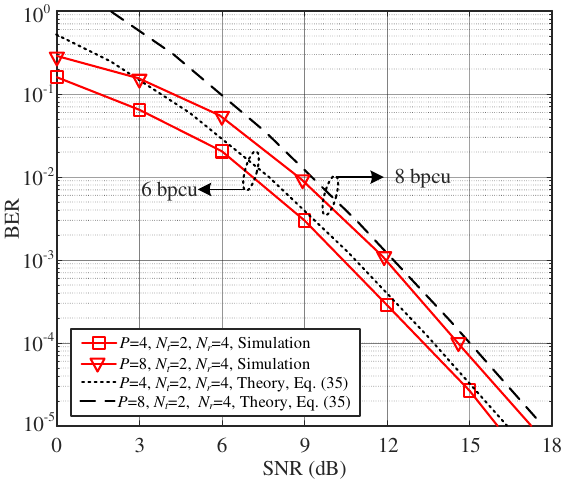}
\caption{Simulation and theoretical results of the proposed FA-IM system using the optimal ML detector under various parameter setting with $L=15$ and 4QAM modulation.}
\label{BER_Theory}
\end{figure}

\begin{figure}[t]
\centering
\includegraphics[width=0.4\textwidth]{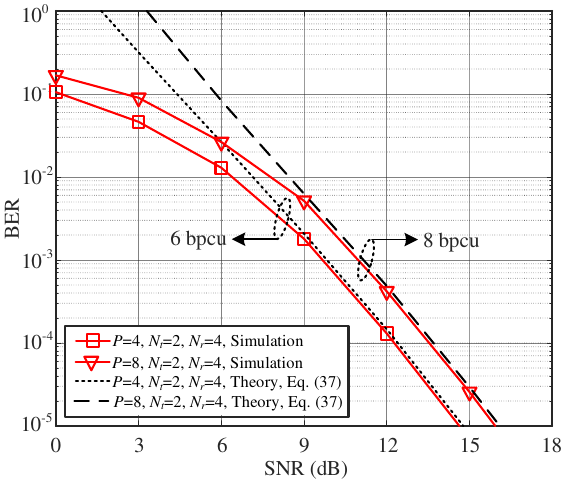}
\caption{Simulation and theoretical results of the proposed FA-IM system using the optimal ML detector under various parameter setting with $L=100$ and 4QAM modulation.}
\label{BER_Theory_2}
\end{figure}

\section{Simulation Results}
In this section, we present the simulation results to evaluate the performance of the proposed FA-IM scheme and validate our derived analytical upper bound for ABEP. Here, we consider the geometry channel model, where the number of the transmit and receive paths are the same, i.e., $L_t=L_r=L$. The elements of the PRM follow the CSCG distribution, i.e.,  ${\bf{\Xi }} = {\rm{diag}}({\alpha _1},...,{\alpha _l},...,{\alpha _L}) \in {{\mathbb {C}}^{L \times L}}$, with ${\alpha _l} \sim {\cal {CN}}(0,c/L)$, where $c=c_0d^{-\varpi}$ is the expected channel gain, $c_0=-30$ dB is the unit distance path loss, $d=25$ m is the distance between the transmit region and the receive region, and $\varpi=2.3$ is the path loss exponent. In addition, the carrier frequency $f$ is 28 GHz, i.e., the wavelength $\lambda$ is 0.01 m, and the transmitter area is a square area of size $10\lambda \times 10\lambda$.

Figs. \ref{BER_Theory} and \ref{BER_Theory_2} evaluate the analytical and simulated BER performance of the proposed FA-IM over the finite-path channel and infinite-path channel, respectively, for different parameter settings using the optimal ML detector. To be specific, the theoretical curves in Figs. \ref{BER_Theory} and \ref{BER_Theory_2} are respectively plotted based on \eqref{UPEP_1} and \eqref{UPEP_11}. It can be observed that the gap between the upper bound and the simulated results becomes more pronounced as the SE increases. This phenomenon stems from the inherent nature of the upper bound, which serves as an approximation, inherently subject to limitations as SE escalates. Nonetheless, the theoretical upper bound exhibits a propensity to approach the simulation result more closely as the SNR increases, thus the upper bound can be served as an effective theoretical tool to evaluate the system performance.

\begin{figure}[t]
\centering
\includegraphics[width=0.42\textwidth]{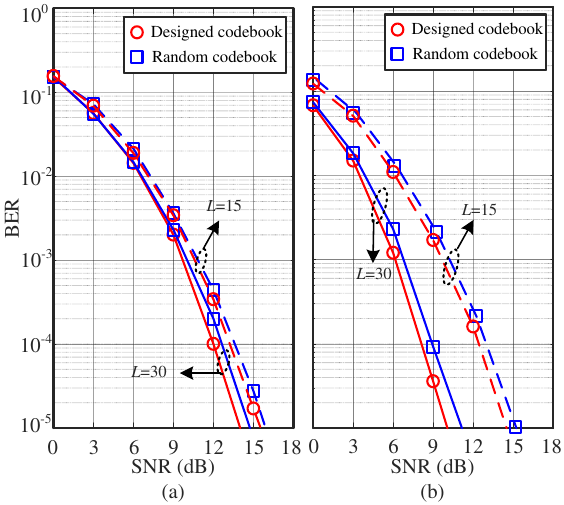}
\caption{BER performance comparison of the proposed FA-IM scheme with designed FA position codebook and random FA position pattern codebook: (a). $P=4,N_t=2,N_r=4$, 4QAM; (b). $P=16,N_t=2,N_r=16$, 8QAM.}
\label{BER_codebook}
\end{figure}

Fig. \ref{BER_codebook} characterizes the BER performance of the proposed FA-IM system with the designed FA position pattern codebook under various parameter settings. As shown in Fig. \ref{BER_codebook}, the designed FA position pattern codebook aided FA-IM system yield a enhancement in BER performance compared to its counterpart employing random codebook. Furthermore, the magnitude of this performance improvement gradually increases with an increase of $A$, where $A$ is the number of possible FA position patterns, as evidenced by the results. Specifically, in the case of $N_t=2$, $P=4$ and $L=30$ (i.e., $A=6$), the designed codebook aided scheme provides 0.6 dB SNR gain compared to the random codebook aided FA-IM scheme at BER = $10^{-5}$. In contrast, it offers 1 dB SNR gain under the condition of $N_t=2$, $P=16$ and $L=30$ (i.e., $A=120$). 

\begin{figure}[t]
\centering
\includegraphics[width=0.4\textwidth]{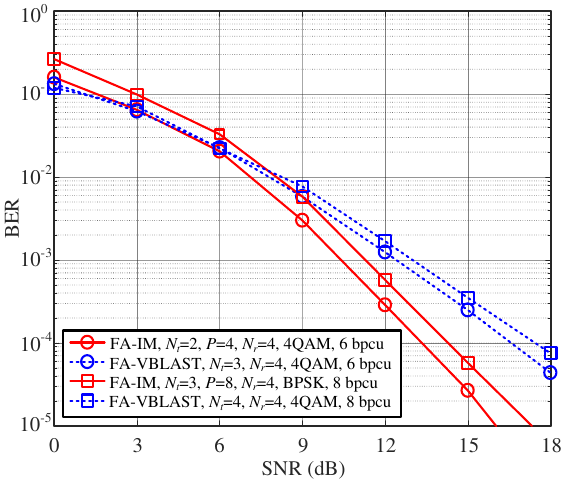}
\caption{BER performance comparisons between the proposed FA-IM and FA-VBLAST schemes under the same transmission rate (i.e., 6 and 8 bpcu).}
\label{BER_diff_system}
\end{figure}

\begin{figure}[t]
\centering
\includegraphics[width=0.42\textwidth]{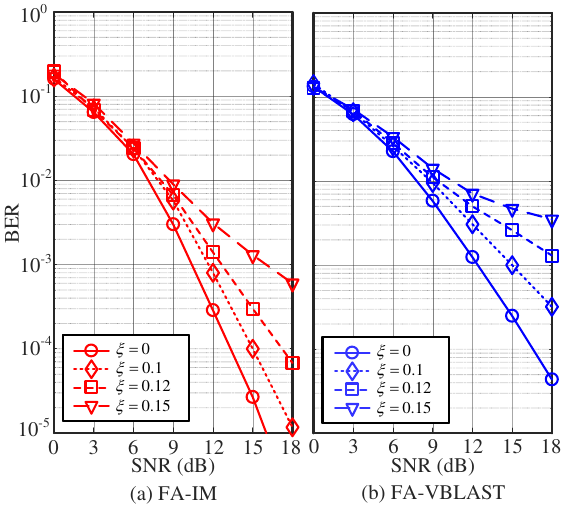}
\caption{BER performance at 6 bpcu for FA-IM and FA-VBLAST schemes in perfect and imperfect CSI cases.}
\label{BER_channel_error}
\end{figure}

\begin{figure}[t]
\centering
\includegraphics[width=0.4\textwidth]{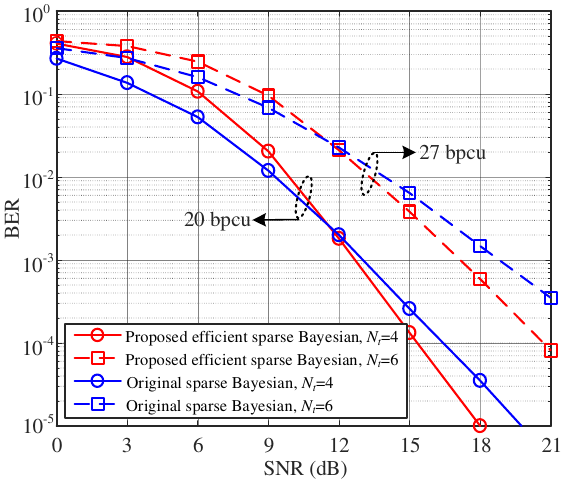}
\caption{BER performance of the proposed FA-IM scheme with the proposed low-complexity detector under different SE configurations.}
\label{BER_Detector1}
\end{figure}

Fig. \ref{BER_diff_system} compares the BER performance of the FA-IM and FA-VBLAST schemes under the same transmission rate{\footnote{The existing literature on FA-assisted MIMO systems mainly focuses on optimizing FA positions to achieve higher capacity or beamforming gains. However, no existing works have considered the BER performance, which reveals the practicality of the schemes. To bridge this gap, we compare the BER performance of the proposed FA-IM and the conventional FA-VBLAST schemes, where the simulation setup strictly follows the work in \cite{zhu2022modeling,zhu2023movable,ma2022mimo}.}}. Both schemes use optimal ML detector. In the case of 8 bpcu, we set the simulation parameter as $N_t=3$, $P=8$, $N_r=4$ and BPSK for the proposed FA-IM scheme, and $N_t=4$, $N_r=4$ and 4QAM for the FA-VBLAST scheme. Compared to the FA-VBLAST scheme, the proposed FA-IM scheme exhibits worse error performance in the low SNR region, this is attributes to the inaccurate estimation of the FA position pattern. However, the proposed FA-IM scheme shows better error performance than FA-VBLAST at ${\rm{SNR}}>8$ dB. Specifically, it can provide nearly 3 dB SNR gain at ${\rm{BER}}=10^{-4}$. This is due to less estimation errors by employing the BPSK modulation.

Fig. \ref{BER_channel_error} examines the effect of channel estimation errors on the BER performance of the FA-IM and FA-VBLAST schemes. To obtain the same transmission rate of 6 bpcu, we set $N_t=2$, $P=N_r=4$ and 4QAM for the proposed FA-IM, and $N_t=3$, $N_r=4$ and 4QAM for the FA-VBLAST scheme. Here, the imperfect channel can be modeled as ${\bf{\hat H}} = \sqrt {1 - {\xi ^2}} {\bf{H}} + \xi \Delta {\bf{H}}$, where $\Delta {\bf{H}}$ obeys the same distribution as $\bf H$, and $\xi$ is the error coefficient. Note that the performance of the perfect CSI scenario (i.e., $\xi=0$) is also included for the purpose of facilitating comparative analysis.  As seen from Fig. \ref{BER_channel_error}, the proposed FA-IM exhibits resilience against channel estimation errors for values of $\xi \le 0.1$. Compared to the FA-VBLAST scheme, the proposed FA-IM scheme exhibits better robustness.  More specifically,
at a BER value of $10^{-3}$, the SNR deterioration of the proposed FA-IM scheme is 1.2 dB for $\xi=0.1$ compared to the ideal CSI case. Notably, for the FA-VBLAST scheme, the degradation increases to 2.5 dB, a disparity that is markedly higher than that observed with FA-IM. Consequently, we can conclude that the proposed FA-IM scheme is more robust to channel estimation errors compared to the FA-VBLAST scheme.

Fig. \ref{BER_Detector1} compares the detection performance between the proposed efficient sparse Bayesian and the original sparse Bayesian detectors. For these two detectors, we set the maximum number of iterations as $T_{\rm{max}}=10$. In addition, the parameter ${\psi _t}$ for pruning in the proposed detector is set to ${{\psi _1}=...={\psi _t}=...={\psi _{{T_b}}}}=2$ and $T_b=5$. Other parameters are set as $(N_t,P,N_r,M)=(4,20,20,4)$ and $(N_t,P,N_r,M)=(6,20,20,4)$  to achieve 20 bpcu and 27 bpcu date rate, respectively. As shown in Fig. \ref{BER_Detector1}, the proposed detector exhibits worse BER performance in the low SNR region compared to the original sparse Bayesian detector. Precisely, during the pruning process, the influence of noise may lead to the discard of elements in the hyperparameter vector that correspond to active FA positions.  As the SNR increases, the pruning process becomes increasingly accurate, thus offering considerable SNR gains over that of the original one. Specifically, the proposed detector provides 1.8 dB SNR gain compared to the original sparse Bayesian detector at BER $=10^{-5}$ when date rate is equal to 20 bpcu. The error performance enhancement comes from the accurate pruning process. To be specific, zeros are the optimal solutions for the inactive FA positions, it means that the accurate pruning process makes the elements in the hyperparameter vector associated with inactive FA positions arrive at the optimal solution in advance. Consequently, we can conclude that the proposed detector not only facilitates  the computational complexity reduction, but also contributes to an enhancement in BER performance as the SNR increases.

\section{Conclusion}
In this paper, we proposed a novel transmission scheme for FAs-assisted MIMO system, namely FA-IM, to achieve SE performance enhancement by exploiting the benefits of the IM technique. Subsequently, a FA position pattern codebook design strategy was presented to enhance the error performance by fully utilizing the effective channel gains. Then, a low-complexity efficient sparse Bayesian detector was proposed, which is amenable to the high SE FA-IM system. Then, the SE performance and the upper bound for ABEP were analyzed. Simulation results demonstrated the superiority of the proposed FA-IM scheme over FA-VBLAST scheme.

\begin{appendices}
\section{Proof of Lemma 1}
According to the estimation of FA position pattern, the variable $Z$ can be divided into two categories
\begin{equation}\label{MA_PP_detection}
Z = \left\{ {\begin{array}{*{20}{l}}
{{{\left\| {\sum\limits_{k = 1}^{{N_t}} {{\bf{h}}({{\bf{t}}_{{\Omega _{{I_k}}}}})({s_k} - {{\hat s}_k})} } \right\|}^2},}&{{\cal I} = {\hat {\cal I}}}\\
{{{\left\| {\sum\limits_{k = 1}^{{N_t}} {\left( {{\bf{h}}({{\bf{t}}_{{\Omega _{{I_k}}}}}){s_k} - {\bf{h}}({{\bf{t}}_{{\Omega _{{{\hat I}_k}}}}}){{\hat s}_k}} \right)} } \right\|}^2},}&{{\cal I} \ne {\hat {\cal I}}}
\end{array}} \right. .
\end{equation}

In the case of ${\cal{I}}={\hat {\cal{I}}}$,  $Z$ can be simplified as
\begin{equation}
\begin{aligned}
Z &= \left| {{s_k} - {{\hat s}_k}} \right|^2{\left\| {\sum\nolimits_{k = 1}^{{N_t}} {\bf{h}}({{\bf{t}}_{{\Omega _{{I_k}}}}})} \right\|^2}\\
& = {\left| {{s_k} - {{\hat s}_k}} \right|^2}{\left\| {\sum\limits_{k = 1}^{{N_t}} {{{\bf{F}}^H}{\bf{\Xi g}}({{\bf{t}}_{{\Omega _{{I_k}}}}})} } \right\|^2}\\
& = {\left| {{s_k} - {{\hat s}_k}} \right|^2}{\left| {\sum\limits_{k = 1}^{{N_t}} {{{\bf{F}}^H}{\bf{g}}({{\bf{t}}_{{\Omega _{{I_k}}}}})} } \right|^2}{\left\| {\bf{\Xi }} \right\|^2}.
\end{aligned}
\end{equation}

In the case of ${\cal{I}}\ne{\hat {\cal{I}}}$,  $Z$ can be represented as
\begin{equation}
\begin{aligned}
Z& = {\left\| {\sum\limits_{k = 1}^{{N_t}} {\left( {{\bf{h}}({{\bf{t}}_{{\Omega _{{I_k}}}}}){s_k} - {\bf{h}}({{\bf{t}}_{{\Omega _{{{\hat I}_k}}}}}){{\hat s}_k}} \right)} } \right\|^2}\\
& = {\left\| {\sum\limits_{k = 1}^{{N_t}} {\left( {{{\bf{F}}^H}{\bf{\Xi g}}({{\bf{t}}_{{\Omega _{{I_k}}}}}){s_k} - {{\bf{F}}^H}{\bf{\Xi g}}({{\bf{t}}_{{\Omega _{{{\hat I}_k}}}}}){{\hat s}_k}} \right)} } \right\|^2}\\
& = {\left| {{{\bf{F}}^H}\sum\limits_{k = 1}^{{N_t}} {\left( {({\bf{g}}({{\bf{t}}_{{\Omega _{{I_k}}}}}){s_k} - {\bf{g}}({{\bf{t}}_{{\Omega _{{{\hat I}_k}}}}}){{\hat s}_k})} \right)} } \right|^2}{\left\| {\bf{\Xi }} \right\|^2}.
\end{aligned}
\end{equation}

Since ${\bf{\Xi }}$ is the PRM, whose elements follow the CSCG distribution, i.e., ${\bf{\Xi }} = {\rm{diag}}({\alpha _1},...,{\alpha _l},...,{\alpha _L}) \in {{\mathbb {C}}^{L \times L}}$, with ${\alpha _l} \sim {\cal {CN}}(0,1/L)$, $Z$ follows the generalized central Chi-square distribution with $2L$ degree of freedom. The MGF of $Z$ is given by
\begin{equation}\label{MGF_1}
{M_Z}(\varepsilon ) = {\left( {1 - \frac{{2L\varepsilon }}{B}} \right)^{ - L}},
\end{equation}
where 
\begin{equation}
B = \left\{ {\begin{array}{*{20}{l}}
{{{\left| {{s_k} - {{\hat s}_k}} \right|}^2}{{\left| {\sum\limits_{k = 1}^{{N_t}} {{{\bf{F}}^H}{\bf{g}}({{\bf{t}}_{{\Omega _{{{\cal I}_k}}}}})} } \right|}^2},}&{{\cal I} = \hat {\cal I}}\\
{\left| {{{\bf{F}}^H}\sum\limits_{k = 1}^{{N_t}} {\left( {({\bf{g}}({{\bf{t}}_{{\Omega _{{{\cal I}_k}}}}}){s_k} - {\bf{g}}({{\bf{t}}_{{\Omega _{{{\hat {\cal I}}_k}}}}}){{\hat s}_k})} \right)} } \right|,}&{{\cal I} \ne \hat {\cal I}}
\end{array}} \right..
\end{equation}
By substituting \eqref{MGF_1} to \eqref{UPEP_3}, we have
\begin{equation}
\begin{aligned}
&{\mathbb{P}}\{ ({\cal{I}},{\bf{s}}) \to (\hat {\cal{I}},{\bf{\hat s}})\}\\
&\approx \frac{1}{{12}}{\left( {1 + \frac{L}{{2{N_0}B}}} \right)^{ - L}} + \frac{1}{4}{\left( {1 + \frac{2L}{{3{N_0}B}}} \right)^{ - L}}.
\end{aligned}
\end{equation}
By taking expectation over enough number of channel realizations, the UPEP expression can be obtained as in \eqref{UPEP_1}.

\section{Proof of Lemma 2}
The channel in \eqref{channel_model} can be rewritten as
\begin{equation}
{\bf{h}}({{\bf{t}}_{{\Omega _{{I_k}}}}}) = {{\bf{F}}^H}{\bf{\Xi g}}({{\bf{t}}_{{\Omega _{{I_k}}}}}) = \sum\limits_{l = 1}^L {\alpha {}_l} {g_l}{\bf{a}}({\theta _{{\rm{r}},l}},{\phi _{{\rm{r}},l}}),
\end{equation}
where ${g_l} = {e^{j\frac{{2\pi }}{\lambda }{\rho _{{\rm{t}},l}}({{\bf{t}}_{{\Omega _{{I_k}}}}})}}$ and ${\bf{a}}({\theta _{{\rm{r}},l}},{\phi _{{\rm{r}},l}})$ is the $l$th row of the FRM $\bf{F}$, denoted as
\begin{equation}
\begin{aligned}
&{\bf{a}}({\theta _{{\rm{r}},l}},{\phi _{{\rm{r}},l}}) = {[{\rho _{{\rm{r}},l}}({{\bf{r}}_{1,1}}),{\rho _{{\rm{r}},l}}({{\bf{r}}_{1,2}}),...,{\rho _{{\rm{r}},l}}({{\bf{r}}_{{N_1},{N_2}}})]^T}\\
& = {[{e^{j{{\bf{k}}_{\rm{r}}}({\theta _{{\rm{r,}}l}},{\phi _{{\rm{r}},l}}){{\bf{r}}_{1,1}}}},{e^{j{{\bf{k}}_{\rm{r}}}({\theta _{{\rm{r,}}l}},{\phi _{{\rm{r}},l}}){{\bf{r}}_{1,2}}}},...,{e^{j{{\bf{k}}_{\rm{r}}}({\theta _{{\rm{r,}}l}},{\phi _{{\rm{r}},l}}){{\bf{r}}_{{N_1},{N_2}}}}}]^T}.       
\end{aligned}
\end{equation}
As $L \to \infty$, it follows the central limit theorem (CLT) that 
\begin{equation}
{\bf{h}}({{\bf{t}}_{{\Omega _{{I_k}}}}}) \sim {\cal N}({\bf{0}},\frac{1}{L}{\bf{R}}), 
\end{equation}
where the normalized spatial correlation matrix ${\bf{R}} \in {\mathbb{C}}^{N_r\times N_r}$ is calculated as
\begin{equation}
{\bf{R}} = L{\mathbb{E}}\{ {\bf{h}}({{\bf{t}}_{{\Omega _{{I_k}}}}}){\bf{h}}{({{\bf{t}}_{{\Omega _{{I_k}}}}})^H}\}  = {\mathbb{E}}\{ {\bf{a}}({\theta _{{\rm{r}},l}},{\phi _{{\rm{r}},l}}){\bf{a}}{({\theta _{{\rm{r}},l}},{\phi _{{\rm{r}},l}})^H}\}.
\end{equation}
The $(i,j)$th element in $\bf{R}$ is given by
\begin{equation}\label{Spatial_correlation_1}
\begin{aligned}
{[{\bf{R}}]_{i,j}} &= {\mathbb{E}}\{ {\bf{a}}({\theta _{{\rm{r}},l}},{\phi _{{\rm{r}},l}})(i){\bf{a}}({\theta _{{\rm{r}},l}},{\phi _{{\rm{r}},l}}){(j)^H}\} \\
& = {\mathbb{E}}\{ {e^{j{{\bf{k}}_{\rm{r}}}({\theta _{{\rm{r,}}l}},{\phi _{{\rm{r}},l}})({{\bf{r}}_{m,n}} - {{\bf{r}}_{p,q}})}}\} \\
& = {\mathbb{E}}\{ {e^{j\frac{{2\pi }}{\lambda }(({x_{{\rm{r}},m}} - {x_{{\rm{r}},p}})\cos {\theta _{{\rm{r}},l}}\sin {\phi _{{\rm{r}},l}} + ({y_{{\rm{r}},n}} - {y_{{\rm{r}},n}})\sin {\theta _{{\rm{r}},l}})}}\}       
\end{aligned},
\end{equation}
where $i,j\in\{1,2,...,N_r\}$ and satisfies the relation with the pairs $(m,n)$ and $(p,q)$ as $i=(m-1)N_1+n$ and $j=(p-1)N_1+q$, $m,p \in \{1,2,...,N_1\}$ and $n,q \in \{1,2,...,N_2\}$.

In isotropic scattering environments, the multi-path components is uniformly distributed over the half-space in front of the UPA, which yields the AoAs following the probability density function (PDF)
\begin{equation}
f({\theta _{{\rm{r}},l}},{\phi _{{\rm{r}},l}}) = \frac{{\cos {\theta _{{\rm{r}},l}}}}{{2\pi }},{\theta _{{\rm{r}},l}} \in [ - \frac{\pi }{2},\frac{\pi }{2}],{\phi _{{\rm{r}},l}} \in [ - \frac{\pi }{2},\frac{\pi }{2}].
\end{equation}

Therefore, the expression in \eqref{Spatial_correlation_1} can be further calculated as
\begin{equation}
\begin{aligned}
{[{\bf{R}}]_{i,j}} &= \int\limits_{ - {\pi  \mathord{\left/
 {\vphantom {\pi  2}} \right.
 \kern-\nulldelimiterspace} 2}}^{{\pi  \mathord{\left/
 {\vphantom {\pi  2}} \right.
 \kern-\nulldelimiterspace} 2}} {\int\limits_{{{ - \pi } \mathord{\left/
 {\vphantom {{ - \pi } 2}} \right.
 \kern-\nulldelimiterspace} 2}}^{{\pi  \mathord{\left/
 {\vphantom {\pi  2}} \right.
 \kern-\nulldelimiterspace} 2}} {{e^{j\frac{{2\pi }}{\lambda }\left\| {{{\bf{r}}_{m,n}} - {{\bf{r}}_{p,q}}} \right\|\sin {\theta _{{\rm{r}},l}}}}} } f({\theta _{{\rm{r}},l}},{\phi _{{\rm{r}},l}}){\rm{d}}{\theta _{{\rm{r}},l}}{\rm{d}}{\phi _{{\rm{r}},l}}\\
& = \int\limits_{{{ - \pi } \mathord{\left/
 {\vphantom {{ - \pi } 2}} \right.
 \kern-\nulldelimiterspace} 2}}^{{\pi  \mathord{\left/
 {\vphantom {\pi  2}} \right.
 \kern-\nulldelimiterspace} 2}} {{e^{j\frac{{2\pi }}{\lambda }\left\| {{{\bf{r}}_{m,n}} - {{\bf{r}}_{p,q}}} \right\|\sin {\theta _{{\rm{r}},l}}}}\frac{{\cos {\theta _{{\rm{r}},l}}}}{{2\pi }}} {\rm{d}}{\theta _{{\rm{r}},l}}\\
& = \frac{{\sin (\frac{{2\pi }}{\lambda }\left\| {{{\bf{r}}_{m,n}} - {{\bf{r}}_{p,q}}} \right\|)}}{{\frac{{2\pi }}{\lambda }\left\| {{{\bf{r}}_{m,n}} - {{\bf{r}}_{p,q}}} \right\|}}.     
\end{aligned}
\end{equation}

The variable $Z$ follows the generalized central Chi-square distribution with $2N_t$ degree of freedom. The MGF of $Z$ is
\begin{equation}\label{MGF_11}
{M_Z}(\varepsilon ) = {\left[ {\det \left( {{{\bf{I}}_{{N_r}}} - \frac{{2L\varepsilon }}{{{\bf{R}}{{\left| {{s_k} - {{\hat s}_k}} \right|}^2}}}} \right)} \right]^{ - N_t}}.
\end{equation}
The UPEP expression can be obtained as in \eqref{UPEP_11} by substituting \eqref{MGF_11} to \eqref{UPEP_3}.

\end{appendices}

\bibliography{ref}

\begin{thebibliography}{10}

\bibitem{bjornson2016massive}
E.~Bj{\"o}rnson, E.~G. Larsson, and T.~L. Marzetta, ``Massive {MIMO}: Ten myths and one critical question,'' {\em IEEE Commun. Mag.}, vol.~54, no.~2, pp.~114--123, Feb. 2016.

\bibitem{albreem2019massive}
M.~A. Albreem, M.~Juntti, and S.~Shahabuddin, ``Massive {MIMO} detection techniques: A survey,'' {\em IEEE Commun. Survey. Tuts}, vol.~21, no.~4, pp.~3109--3132, 4th Quart., 2019.

\bibitem{wang2019overview}
M.~Wang, F.~Gao, S.~Jin, and H.~Lin, ``An overview of enhanced massive {MIMO} with array signal processing techniques,'' {\em IEEE J. Sel. Top. Sign. Process.}, vol.~13, no.~5, pp.~886--901, Sep. 2019.

\bibitem{wong2020fluid}
K.-K. Wong, A.~Shojaeifard, K.-F. Tong, and Y.~Zhang, ``Fluid antenna systems,'' {\em IEEE Trans. Wireless Commun.}, vol.~20, no.~3, pp.~1950--1962, Mar. 2021.

\bibitem{wong2021fluid}
K.-K. Wong and K.-F. Tong, ``Fluid antenna multiple access,'' {\em IEEE Trans. Wireless Commun.}, vol.~21, no.~7, pp.~4801--4815, Jul. 2022.

\bibitem{wong2022bruce}
K.-K. Wong, K.-F. Tong, Y.~Shen, Y.~Chen, and Y.~Zhang, ``Bruce {L}ee-inspired fluid antenna system: Six research topics and the potentials for {6G},'' {\em Frontiers Commun. Netw.}, vol.~3, p.~853416, Mar. 2022.

\bibitem{shen2022radiation}
Y.~Shen, K.-F. Tong, and K.-K. Wong, ``Radiation pattern diversified double-fluid-channel surface-wave antenna for mobile communications,'' in {\em 2022 {IEEE-APS} Topical Conference on Antennas and Propagation in Wireless Communications (APWC)}, pp.~85--88, IEEE, Cape Town, South Africa, Sep. 2022.

\bibitem{zhu2022modeling}
L.~Zhu, W.~Ma, and R.~Zhang, ``Modeling and performance analysis for movable antenna enabled wireless communications,'' {\em arXiv preprint arXiv:2210.05325}, 2022.

\bibitem{beyene2017performance}
Y.~D. Beyene, R.~Jantti, K.~Ruttik, and S.~Iraji, ``On the performance of narrow-band internet of things {(NB-IoT)},'' in {\em IEEE Wireless Commun. Networking Conf. (WCNC)}, pp.~1--6, IEEE, Mar. 2017.

\bibitem{dowhuszko2016performance}
A.~A. Dowhuszko, G.~Corral-Briones, J.~H{\"a}m{\"a}l{\"a}inen, and R.~Wichman, ``Performance of quantized random beamforming in delay-tolerant machine-type communication,'' {\em IEEE Trans. Wireless Commun.}, vol.~15, no.~8, pp.~5664--5680, Aug. 2016.

\bibitem{ma2022mimo}
W.~Ma, L.~Zhu, and R.~Zhang, ``{MIMO} capacity characterization for movable antenna systems,'' {\em arXiv preprint arXiv:2210.05396}, 2022.

\bibitem{zhu2023movable}
L.~Zhu, W.~Ma, and R.~Zhang, ``Movable antennas for wireless communication: opportunities and challenges,'' {\em arXiv preprint arXiv:2306.02331}, 2023.

\bibitem{ma2023compressed}
W.~Ma, L.~Zhu, and R.~Zhang, ``Compressed sensing based channel estimation for movable antenna communications,'' {\em arXiv preprint arXiv:2306.04333}, 2023.

\bibitem{zhuravlev2015experimental}
A.~Zhuravlev, V.~Razevig, S.~Ivashov, A.~Bugaev, and M.~Chizh, ``Experimental simulation of multi-static radar with a pair of separated movable antennas,'' in {\em Proc. IEEE COMCAS}, pp.~1--5, Tel Aviv, Israel, Nov 2015.

\bibitem{basbug2017design}
S.~Basbug, ``Design and synthesis of antenna array with movable elements along semicircular paths,'' {\em IEEE Antennas and Wireless Propag. Lett.}, vol.~16, pp.~3059--3062, Oct. 2017.

\bibitem{wu2023movable}
Y.~Wu, D.~Xu, D.~W.~K. Ng, W.~Gerstacker, and R.~Schober, ``Movable antenna-enhanced multiuser communication: Optimal discrete antenna positioning and beamforming,'' {\em arXiv preprint arXiv:2308.02304}, 2023.

\bibitem{dang2017adaptive}
S.~Dang, J.~P. Coon, and G.~Chen, ``Adaptive {OFDM} with index modulation for two-hop relay-assisted networks,'' {\em IEEE Trans. Wireless Commun.}, vol.~17, no.~3, pp.~1923--1936, Mar. 2018.

\bibitem{zhu2022index}
J.~Zhu, P.~Gao, G.~Chen, P.~Xiao, and A.~Quddus, ``Index modulation for {STAR-RIS} assisted {NOMA} system,'' {\em IEEE Commun. Lett}, vol.~27, no.~2, pp.~716--720, Feb. 2023.

\bibitem{dang2018lexicographic}
S.~Dang, G.~Chen, and J.~P. Coon, ``Lexicographic codebook design for {OFDM} with index modulation,'' {\em IEEE Trans. Wireless Commun.}, vol.~17, no.~12, pp.~8373--8387, Dec. 2018.

\bibitem{wen2019index}
M.~Wen, X.~Chen, Q.~Li, E.~Basar, Y.-C. Wu, and W.~Zhang, ``Index modulation aided subcarrier mapping for dual-hop {OFDM} relaying,'' {\em IEEE Trans. Commun.}, vol.~67, no.~9, pp.~6012--6024, Sep. 2019.

\bibitem{wen2021joint}
M.~Wen, J.~Li, S.~Dang, Q.~Li, S.~Mumtaz, and H.~Arslan, ``Joint-mapping orthogonal frequency division multiplexing with subcarrier number modulation,'' {\em IEEE Trans. Commun.}, vol.~69, no.~7, pp.~4306--4318, Jul. 2021.

\bibitem{mesleh2008spatial}
R.~Y. Mesleh, H.~Haas, S.~Sinanovic, C.~W. Ahn, and S.~Yun, ``Spatial modulation,'' {\em IEEE Trans. Veh. Technol.}, vol.~57, no.~4, pp.~2228--2241, Jul. 2008.

\bibitem{xiao2014low}
Y.~Xiao, Z.~Yang, L.~Dan, P.~Yang, L.~Yin, and W.~Xiang, ``Low-complexity signal detection for generalized spatial modulation,'' {\em IEEE Commun. Lett}, vol.~18, no.~3, pp.~403--406, Mar. 2014.

\bibitem{li2016generalized}
J.~Li, M.~Wen, X.~Cheng, Y.~Yan, S.~Song, and M.~H. Lee, ``Generalized precoding-aided quadrature spatial modulation,'' {\em IEEE Trans. Veh. Technol.}, vol.~66, no.~2, pp.~1881--1886, Feb. 2017.

\bibitem{yang2019enhanced}
P.~Yang, Y.~Xiao, M.~Xiao, J.~Zhu, S.~Li, and W.~Xiang, ``Enhanced receive spatial modulation based on power allocation,'' {\em IEEE J. Sel. Top Signal Process.}, vol.~13, no.~6, pp.~1312--1325, Oct. 2019.

\bibitem{zhu2018low}
J.~Zhu, P.~Yang, Y.~Xiao, X.~Lei, and Q.~Chen, ``Low {RF}-complexity receive spatial modulation for millimeter-wave {MIMO} communications,'' {\em IEEE Commun. Lett.}, vol.~22, no.~7, pp.~1338--1341, Jul. 2018.

\bibitem{gao2019spatial}
S.~Gao, X.~Cheng, and L.~Yang, ``Spatial multiplexing with limited {RF} chains: Generalized beamspace modulation {(GBM)} for mmwave massive {MIMO},'' {\em IEEE J. Sel. Areas Commun.}, vol.~37, no.~9, pp.~2029--2039, Sep. 2019.

\bibitem{ding2017spatial}
Y.~Ding, K.~J. Kim, T.~Koike-Akino, M.~Pajovic, P.~Wang, and P.~Orlik, ``Spatial scattering modulation for uplink millimeter-wave systems,'' {\em IEEE Commun. Lett}, vol.~21, no.~7, pp.~1493--1496, Jul. 2017.

\bibitem{sanayei2004antenna}
S.~Sanayei and A.~Nosratinia, ``Antenna selection in {MIMO} systems,'' {\em IEEE Commun. mag.}, vol.~42, no.~10, pp.~68--73, Oct. 2004.

\bibitem{balanis2008mems}
C.~A. Balanis, ``Mems integrated and micromachined antenna elements, arrays, and feeding networks,'' 2008.

\bibitem{rajashekar2013antenna}
R.~Rajashekar, K.~Hari, and L.~Hanzo, ``Antenna selection in spatial modulation systems,'' {\em IEEE Commun. Lett.}, vol.~17, no.~3, pp.~521--524, Mar. 2013.

\bibitem{ji2008bayesian}
S.~Ji, Y.~Xue, and L.~Carin, ``Bayesian compressive sensing,'' {\em IEEE Trans. signal process.}, vol.~56, no.~6, pp.~2346--2356, Jun. 2008.

\bibitem{al2017gamp}
M.~Al-Shoukairi, P.~Schniter, and B.~D. Rao, ``A {GAMP}-based low complexity sparse bayesian learning algorithm,'' {\em IEEE Trans. Signal Process.}, vol.~66, no.~2, pp.~294--308, Jan. 2018.

\bibitem{wipf2007empirical}
D.~P. Wipf and B.~D. Rao, ``An empirical bayesian strategy for solving the simultaneous sparse approximation problem,'' {\em IEEE Trans. Signal Process.}, vol.~55, no.~7, pp.~3704--3716, Jul. 2007.

\bibitem{zhang2022low}
X.~Zhang, H.~Zeng, B.~Ji, and G.~Zhang, ``Low-complexity implicit detection for massive {MIMO} using neumann series,'' {\em IEEE Trans. Veh. Technol.}, vol.~71, no.~8, pp.~9044--9049, Aug. 2022.

\bibitem{simon2001digital}
M.~K. Simon and M.-S. Alouini, {\em Digital communication over fading channels}.
\newblock New York: Wiley, 2001.

\bibitem{ma2020large}
T.~Ma, Y.~Xiao, X.~Lei, P.~Yang, X.~Lei, and O.~A. Dobre, ``Large intelligent surface assisted wireless communications with spatial modulation and antenna selection,'' {\em IEEE J. Sel. Areas Commun.}, vol.~38, no.~11, pp.~2562--2574, Nov. 2020.

\end{thebibliography}
\bibliographystyle{ieeetr}

\end{document}